\documentclass{elsart3p}

\usepackage{natbib}

\usepackage{graphicx}

\usepackage{amssymb}

\begin{document}

\def\lta{\lower2pt\hbox{$\buildrel {\scriptstyle <}
   \over {\scriptstyle\sim}$}}
\def\gta{\lower2pt\hbox{$\buildrel {\scriptstyle >}
   \over {\scriptstyle\sim}$}}
\def\msun{M_\odot}

\begin{frontmatter}


\title{Advection-Dominated Accretion and the Black Hole Event Horizon}
\author{Ramesh Narayan}, 
\ead{rnarayan@cfa.harvard.edu}
\author{Jeffrey E. McClintock}
\ead{jmcclintock@cfa.harvard.edu}
\address{Harvard-Smithsonian Center for Astrophysics, Cambridge, MA 02138, U.S.A.}

\begin{abstract}
As the luminosity of an accreting black hole drops to a few percent of
Eddington, the spectrum switches from the familiar soft state to a
hard state that is well-described by a distended and tenuous
advection-dominated accretion flow (ADAF).  An ADAF is a poor
radiator, and the ion temperature can approach $10^{12}$~K near the
center, although the electrons are cooler, with their temperature
typically capped at $\sim 10^{9-11}$~K.  The foundational papers
predicted that the large thermal energy in an ADAF would drive strong
winds and jets, as later observed and also confirmed in computer
simulations.  Of chief interest, however, is the accreting gas that
races inward.  It carries the bulk of the accretion energy as stored
thermal energy, which vanishes without a trace as the gas passes
through the hole's event horizon.  One thus expects black holes in the
ADAF regime to be unusually faint.  Indeed, this is confirmed by a
comparison of accreting stellar-mass black holes and neutron stars,
which reside in very similar transient X-ray binary systems.  The
black holes are on average observed to be fainter by a factor of $\sim
100-1000$.  The natural explanation is that a neutron star must
radiate the advected thermal energy from its surface, whereas a black
hole can hide the energy behind its event horizon.  The case for an
event horizon in Sagittarius A$^*$, which is immune to caveats on jet
outflows and is furthermore independent of the ADAF model, is
especially compelling.  These two lines of evidence for event horizons
are impervious to counterarguments that invoke strong gravity or
exotic stars.
\end{abstract}

\begin{keyword}


\end{keyword}

\end{frontmatter}

\section{Introduction}
\label{Intro}

The foundations of our present understanding of {\it
advection-dominated accretion} were laid out in a series of papers by
Narayan \& Yi (1994, 1995a,b, hereafter NY94, NY95a, NY95b),
Abramowicz et al. (1995) and Chen et al. (1995), although some ideas
were anticipated much earlier by Ichimaru (1977).  The specific
abbreviation ADAF, which stands for {\it advection-dominated accretion
flow}, was introduced by Lasota\footnote{It gives the authors great
pleasure to highlight Jean-Pierre Lasota's research on ADAFs in this
contribution to his Festschrift.}(1996) and has become standard in the
field.  We review here the application of ADAFs to accreting black
holes (BHs), and in \S4 we also consider their application to
accreting neutron stars.

\section{Accretion Regimes}

The energy equation per unit volume of an accretion flow can be
written compactly in the form
\begin{equation}
\rho T{dS\over dt} \equiv 
\rho T \left({\partial S\over \partial t} +
\vec{v}\cdot\vec{\nabla}S\right) = q_+ - q_-,
\end{equation}
where $\rho$ is the density, $T$ is the temperature, $S$ is the
entropy per unit mass, $t$ is time, $\vec{v}$ is the flow velocity,
and $q_+$ and $q_-$ are the heating and cooling rates per unit volume.
This equation states that the rate at which the entropy per unit
volume of the gas increases is equal to the heating rate minus the
cooling rate.  Since any entropy stored in the gas is advected with
the flow, the left-hand side of equation (1) may be viewed as the
effective {\it advective cooling} rate $q_{\rm adv}$.  Equation (1)
can then be rearranged to give
\begin{equation}
q_+ = q_- + q_{\rm adv},
\end{equation}
which states that the heat energy released by viscous dissipation is
partially lost by radiative cooling and partially by advective
cooling.

The standard thin accretion disk model (Shakura \& Sunyaev 1973;
Novikov \& Thorne 1973; Frank, King \& Raine 2002) corresponds to the
case when the accreting gas is {\it radiatively efficient}, so that we have
\begin{equation}
{\rm Thin~Disk}: \quad q_- \gg q_{\rm adv}, \quad L \sim 0.1\dot{M}c^2.
\end{equation}  
Since the gas cools efficiently, the sound speed is much less than the
local Keplerian speed $v_K$ and the disk is geometrically thin.  Also,
the disk radiates about a tenth of the rest mass energy of the
accreting gas, the precise fraction depending on the BH spin (see
Shapiro \& Teukolsky 1983).

An ADAF corresponds to the opposite regime.  Here the gas is {\it
radiatively inefficient} and the accretion flow is underluminous.
Thus, an ADAF is defined by the condition
\begin{equation}
{\rm ADAF}: \quad q_{\rm adv} \gg q_-, \quad L \ll 0.1\dot{M}c^2.
\label{ADAFdef}
\end{equation}
Some papers in the literature define an ADAF as a flow that
corresponds exactly to a self-similar solution described in NY94 (see
\S3.1).  This is a needless restriction.  In our opinion, it is more
fruitful to employ the general definition of an ADAF as given in
equation (\ref{ADAFdef})\footnote{Actually, at mass accretion rates
for which the flow is close to the boundary between the thin disk and
ADAF solutions, we have (by continuity) $q_{\rm adv} \ \gta\ q_-$, and
the flow is only marginally radiatively inefficient (see Fig. 4).
Nevertheless, even here, the two solutions are very distinct from each
other.}.

There are two distinct regimes of advection-dominated accretion.  The
first, the one that we focus on in this article is when the accreting
gas is very tenuous and has a long cooling time (NY94; NY95b;
Abramowicz et al. 1995).  This regime is sometimes referred to as a
RIAF --- a ``radiatively inefficient accretion flow'' --- and is
defined by the condition
\begin{equation}
{\rm ADAF/RIAF}: \quad t_{\rm cool} \gg t_{\rm acc}.
\end{equation}
Here $t_{\rm cool}$ is the cooling time of the gas and $t_{\rm acc}$
is the accretion time.

The second regime of advection-dominated accretion is when the
particles in the gas have no trouble cooling, but the scattering
optical depth of the accretion flow is so large that the radiation is
unable to diffuse out of the system.  This radiation-trapped regime
was briefly discussed by Begelman (1979) and was developed in detail
by Abramowicz et al. (1988) in their ``slim disk'' model.  The
defining condition for this regime of ADAFs is
\begin{equation}
{\rm ADAF/Slim~Disk}: \quad t_{\rm diff} \gg t_{\rm acc},
\end{equation}
where $t_{\rm diff}$ is the diffusion time for photons.

The present review is devoted exclusively to the ADAF/RIAF form of
accretion (see Narayan, Mahadevan \& Quataert 1998b; Kato, Fukue \&
Mineshige 1998; Lasota 1999a,b; Quataert 2001; Narayan 2002, 2005;
Igumenshchev 2004; Done, Gierlinsky \& Kubota 2007 for reviews
emphasizing various aspects of ADAFs).  We will henceforth drop the
modifier RIAF and refer to these flows simply as ADAFs.

As explained above, an ADAF is by definition very different from the
standard thin accretion disk.  Correspondingly, it is characterized by
very distinct observational signatures.  Both ADAFs and thin disks
have well-known counterparts in nature.

\section{Properties of ADAFs}

\subsection{Geometry and Kinematics}

In an ADAF, since most of the energy released by viscous dissipation
is retained in the gas, the pressure is large, and so is the sound
speed: $c_s \sim v_K$, where the Keplerian velocity $v_K(R)$ is equal
to $c/\sqrt{2r}$ with $r=R/R_S$ being the radius in Schwarzschild
units, $R_S=2GM/c^2$.  The large pressure has several immediate
consequences.  First, the accretion flow becomes geometrically thick,
with a vertical height $H$ of order the radius $R$ (an ADAF may be
viewed as the viscous rotating analog of spherical Bondi accretion).
Second, the flow has considerable pressure-support in the radial
direction, so the angular velocity becomes sub-Keplerian.  Third, the
radial velocity of the gas is relatively large: $v \sim \alpha
v_K(H/R)^2 \sim \alpha c/r^{1/2}$, where $\alpha\sim0.1-0.3$ (see
below) is the standard dimensionless viscosity parameter (Shakura \&
Sunyaev 1973).  Fourth, the large radial velocity leads to a short
accretion time: $t_{\rm acc} =R/v \sim t_{\rm ff}/\alpha$, where
$t_{\rm ff}=(2GM/R^3)^{1/2}$ is the free-fall time.  Finally, the
large velocity and large scale height cause the gas density to be very
low, and so the cooling time is very long and the medium is optically
thin.

The above properties are nicely illustrated in the self-similar ADAF
solution derived by NY94, in which all quantities in the accretion
flow behave as power-laws in radius.  (A similar solution was obtained
by Spruit et al. 1987 in a different context.)  NY94 obtained a
general solution for arbitrary viscosity parameter $\alpha$, adiabatic
index $\gamma$, and advection parameter\footnote{By this definition,
$f= 0$ corresponds to a fully cooling-dominated (no advection) flow
and $f=1 $ corresponds to a fully advection-dominated (no radiative
cooling) flow.} $f \equiv q_{\rm adv}/q_+$.  In the limit
$\alpha^2\ll1$ (a good approximation) and $f\to1$ (radiatively very
inefficient flow), the solution simplifies to
\begin{eqnarray}
v &=& -\alpha \left[{\gamma-1\over\gamma-5/9}\right]v_K = 
-0.53 \alpha v_K, \\
\Omega &=& \left[{2(5/3-\gamma)\over3(\gamma-5/9)}\right]^{1/2}\Omega_K = 
0.34 \Omega_K, \\
c_s &=& \left[{2(\gamma-1)\over3(\gamma-5/9)}\right]^{1/2}v_K = 
0.59 v_K.
\end{eqnarray}
The numerical coefficients on the right correspond to $\gamma=1.5$, a
reasonable choice (see Quataert \& Narayan 1999).

The great virtue of the above self-similar solution is that it is
analytic and provides an easy and transparent way of understanding all
the key properties of an ADAF.  Its biggest deficiency is that it is
scale-free, which means that it is inappropriate near the inner or
outer boundary of the flow.  Therefore, for detailed work, one must
use global solutions of the ADAF equations that satisfy appropriate
boundary conditions (e.g., Abramowicz et al. 1996; Narayan, Kato \&
Honma 1997c; Chen, Abramowicz \& Lasota 1997; Manmoto, Mineshige \&
Kusunose 1997; Popham \& Gammie 1998; Manmoto 2000).  Calculating
global solutions is somewhat involved and one may wish to employ some
short-cuts (e.g., Yuan, Ma \& Narayan 2008).  On the other hand, even
the exact global solutions are somewhat limited since they solve a set
of height-integrated equations.  For greater realism, one might wish
to work directly with numerical simulations (e.g., Goldston, Quataert
\& Igumenshchev 2005; Noble et al. 2007).

\subsection{Thermal Properties}

The ADAF solution is gas pressure dominated.  Since $c_s\sim v_K$,
this means the gas temperature is nearly virial.  Under normal
conditions, gas at such a high temperature will radiate copiously,
especially at small radii where the temperature can approach $10^{12}$
K.  Thus, in order to have an ADAF, the accreting gas generally has to
be a {\it two-temperature plasma} (at least at small radii), with
electron temperature $T_e$ much less than the ion temperature $T_i$
(NY95b).  (The only way to avoid this condition is by having an
extremely low accretion rate below about $10^{-6}$ of the Eddington
rate.)  Typical ADAF models have the two temperatures scaling as
\begin{equation}
T_i \sim 10^{12}{\rm K}/r, \quad T_e \sim
{\rm Min}\, (T_i,\, 10^{9-11} {\rm K}).
\end{equation}

In order for gas in an ADAF to be two-temperature, there must be weak
coupling between electrons and ions.  Models generally assume that the
coupling occurs via Coulomb collisions, which is inefficient at the
densities under consideration.  Begelman \& Chiueh (1988) investigated
whether plasma instabilities might enhance the coupling and drive the
plasma rapidly to a single temperature; this would be problematic for
the ADAF solution.  However, they were unable to identify a clear
mechanism.  For normal mass accretion rates, the electrons in a
two-temperature ADAF will have a thermal energy distribution (but not
necessarily the ions, see Mahadevan \& Quataert 1997).

Early work on two-temperature ADAFs assumed that viscous heating acts
primarily on the ions; for instance, NY95b took the ratio of electron
heating to total heating, $\delta$, to be zero, while Esin, McClintock
\& Narayan (1997) assumed $\delta \sim m_e/m_p \sim10^{-3}$.  However,
an ADAF does not {\it require} $\delta$ to be this small.  Because of
a degeneracy in model parameters (Quataert \& Narayan 1999), it is
possible to have a viable ADAF model with larger values of $\delta$,
provided the mass loss parameter $s$, defined in \S3.6, is adjusted.
More recent ADAF models (e.g., Yuan, Quataert \& Narayan 2003)
typically assume $\delta\sim0.3-0.5$.

Various attempts have been made to estimate the value of $\delta$ from
first principles by considering the effects of reconnection
(Bisnovatyi-Kogan \& Lovelace 1997; Quataert \& Gruzinov 1999) or MHD
turbulence (Quataert 1998; Blackman 1999; Medvedev 2000).  These
studies do not agree on a single value of $\delta$, but generally
suggest that $\delta$ is likely to be much larger than $10^{-3}$.
Recently, Sharma et al. (2007) considered heating by the dissipation
of pressure anisotropy and showed that $\delta\sim (T_e/T_i)^{1/2}/3$.

It should be noted that, even if electrons and ions receive equal
amounts of the dissipated energy (i.e., $\delta=0.5$), the plasma can
still be two-temperature.  This is because a large part of the heating
in an ADAF is by compression (since the density increases inward).
Once $kT \ \gta\ m_e c^2$, which is the case for $r\ \lta$ a few
hundred, the electrons become relativistic and have an adiabatic index
$\gamma_e\sim4/3$, whereas the ions continue to be non-relativistic
with $\gamma_i\sim5/3$.  Since adiabatic heating by compression causes
the temperature to scale as $T \propto \rho^{\gamma-1}$, the electrons
heat up only as $T_e \sim \rho^{1/3}$ whereas the ions heat up as $T_i
\sim \rho^{2/3}$.  Thus, even in the limiting case of $\delta\sim0.5$,
ADAFs naturally become two-temperature at small radii.  For instance,
in the ADAF model of Sagittarius A$^*$ (Sgr A$^*$) proposed by Yuan et
al. (2003), the authors obtain $T_e \sim 0.1T_i$ close to the BH even
though they assumed $\delta=0.55$.

An important property of the ADAF solution is that it is thermally
stable (NY95b; Wu \& Li 1996; Kato et al. 1997; Wu 1997).  The
demonstration of this property was a crucial advance.  Two decades
earlier, in a seminal paper, Shapiro, Lightman \& Eardley (1976), and
after them Rees et al. (1982; 'ion tori'), introduced the idea of a
two-temperature accretion flow and derived a hot two-temperature
accretion flow solution, the SLE solution.  However, that solution
turned out to be thermally unstable (Piran 1978).  Until the
development of the ADAF solution, and the recognition that it is
different from the SLE solution, no stable, hot, optically thin
solution was available to model the many accretion systems whose
spectra (especially in the hard state, see below) demand such a flow.

Chen et al. (1995) and Yuan (2003) have explored the relationships
among the ADAF/RIAF, SLE, ADAF/Slim Disk and Thin Disk solutions.

\subsection{Models and Spectra}

The ADAF is a full accretion solution which incorporates consistent
dynamics, thermal balance, radiation physics, etc.  Therefore, the
radial profiles of all gas properties can be calculated
self-consistently once we know the values of certain parameters: BH
mass $M$, accretion rate $\dot{M}$, viscosity parameter $\alpha$,
pressure parameter $\beta\equiv P_{\rm gas}/(P_{\rm gas}+P_{\rm
mag})$, adiabatic index $\gamma$ (usually $\sim1.5$), viscous heating
parameter $\delta$, advection parameter $f$.

Actually, apart from the system-specific parameters $M$ and $\dot{M}$
(which may be estimated through observation), most of the other
parameters are constrained.  Under the nearly collisionless conditions
expected in an ADAF, the viscosity parameter is moderately enhanced
relative to a collisional gas (Sharma et al. 2006).  In the case of
dwarf novae in the hot state, Smak (1999) estimates $\alpha\sim0.2$,
while numerical simulations of the magneto-rotational instability give
$\alpha\ \lta\ 0.1$ (Hawley, Gammie \& Balbus 1996).  Thus we expect
$\alpha\sim0.1-0.3$ for an ADAF (typical values used in models are
$0.2-0.3$).  Numerical simulations further suggest that magnetic
fields are generally subthermal, with $P_{\rm mag}/P_{\rm gas}\sim
0.1$ (e.g., Hawley et al. 1996), so we expect $\beta\sim0.9$.  By
calculating the energy loss via radiation from the hot accretion flow
(synchrotron, bremsstrahlung, Compton scattering), $f$ can be obtained
self-consistently (e.g., NY95b; Narayan, Barret \& McClintock 1997a;
Esin et al. 1997; Yuan et al. 2003).  Thus, we have only one poorly
constrained parameter: $\delta$.  The recent work of Sharma et
al. (2007) provides a serviceable prescription even for this
parameter.  At large radii, where the plasma is effectively
one-temperature, their formula gives $\delta\sim0.3$, while in the
energetically important inner region, where the plasma is
two-temperature, they find $\delta\sim 0.01-0.1$, depending on model
details.  The final two parameters are discussed later: the transition
radius $R_{\rm tr}$, \S3.5, and the wind parameter $s$, \S3.6.

\begin{figure}
\includegraphics[width=3.3in,clip]{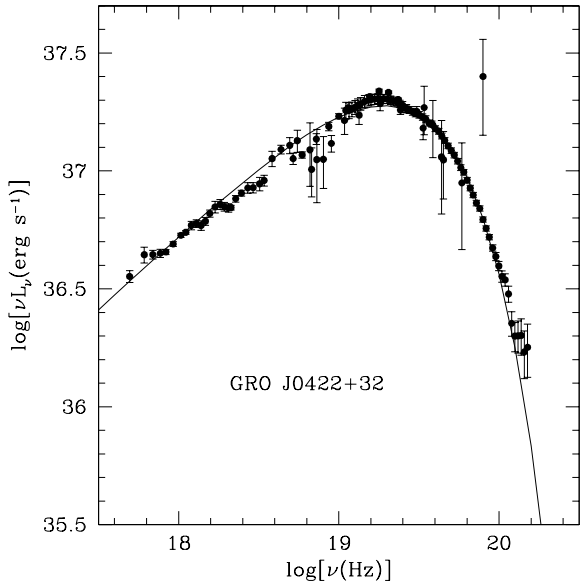}

\caption{Combined TTM (2--20 keV), HEXE (20--200 keV), and OSSE
(50--600 keV) spectrum of GRO J0422+32 observed between 1992 August 29
and September 2.  The solid line is an ADAF model which reproduces the
overall luminosity of the source, the power-law slope below the peak
and the shape of the high-energy cutoff.  The model has a mass
accretion rate $\sim 0.1\dot{M}_{\rm Edd}$. (From Esin et al. 1998)}
\end{figure}

Since the earliest days of X-ray astronomy, it has been clear that BH
binaries (BHBs) have a number of distinct spectral states (see
Zdziarski \& Gierlinski 2004; McClintock \& Remillard 2006; Done et
al. 2007).
The most notable among these are the luminous {\it high soft state},
or {\it thermal state}, the slightly less luminous {\it low hard
state}, and the very under-luminous {\it quiescent state}.

The thermal state is well described by the thin disk model, and a
multi-color disk (MCD) blackbody model (e.g., {\it diskbb}, Mitsuda et
al. 1984; {\it ezdiskbb}, Zimmerman et al. 2005; both available in
XSPEC, Arnaud et al. 1996) has been successfully used for years to
model the X-ray spectra of sources in this state.  Recently, fully
relativistic versions of the MCD model for arbitrary BH spin ({\it
kerrbb}, Li et al. 2005; {\it bhspec}, Davis \& Hubeny 2006) have been
developed, based on the relativistic thin disk model of Novikov \&
Thorne (1973).  These models provide excellent fits to the X-ray
spectra of BHBs in the thermal state (e.g., McClintock et al. 2006;
Davis, Done \& Blaes 2006).

Whereas a satisfactory theoretical model, viz., the thin disk model,
was established early on for the thermal state, the hard state was for
many years a mystery.  Figure 1 shows the spectrum of a typical BHB,
GRO J0422+32, in the hard state (Esin et al. 1998).  The observations
indicate that the accreting gas is very hot, $kT\ \gta\ 100$ keV.  The
gas must also be optically thin, since optically thick blackbody
emission at a temperature of 100 keV would correspond to a luminosity
$L=\sigma T^4 A\ \gta\ 10^{46} ~{\rm erg\,s^{-1}}$ for any reasonable
estimate of the radiating area $A$.

The most natural explanation of the emission in J0422 and other BHBs
in the hard state is that it is produced by thermal Comptonization.
Until the ADAF model was established, no accretion model could
reproduce such a spectrum.  (The SLE solution could, but it was
unstable.)  Indeed, astronomers were reduced to using empirical
Comptonization models (Sunyaev \& Titarchuk 1980) in which they
postulated a Comptonizing cloud with some arbitrary geometry and
parameterized the cloud with an adjustable temperature and an optical
depth (e.g., Zdziarski et al. 1996, 1998; Gierlinski et al. 1997).

The situation changed with the recognition of the ADAF solution.  This
model turned out to have the precise properties --- density, electron
temperature, stability --- needed to explain the hard state.  The
solid line in Fig. 1 shows an ADAF model of J0422 (Esin et al. 1998)
in which the accretion rate has been adjusted to fit the spectrum; the
required rate is about a tenth of the Eddington mass accretion rate
$\dot{M}_{\rm Edd}$, where $\dot{M}_{\rm Edd}=L_{\rm Edd}/(0.1c^2)$,
i.e., it is the mass accretion rate at which a disk with radiative
efficiency 0.1 would radiate at the Eddington luminosity.  It is
gratifying that both the temperature (which determines the position of
the peak) and the Compton $y$-parameter (which determines the
power-law slope below the peak) are reproduced well.  Esin et
al. (1997, 1998, 2001) present models of other BHBs in the hard state.

\begin{figure}
\vskip-0.2in
\includegraphics[width=3.7in,clip]{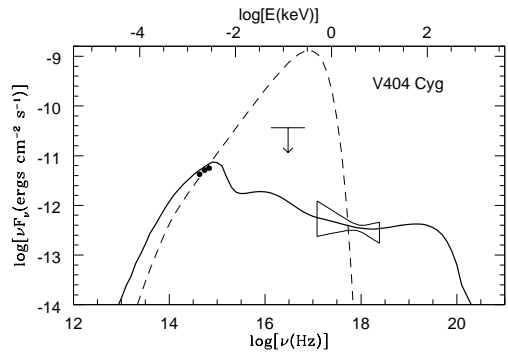}
\vskip -1.5in
\caption{The spectrum of the BHB source V404 Cyg in quiescence.  The
three dots on the left represent optical fluxes (Narayan et al. 1996),
the error box in the X-ray band corresponds to $2\sigma$ limits from
ASCA (Narayan et al. 1997a), and the upper limit in the EUV is derived
from the absence of a HeII $\lambda4686$ line.  The solid line
corresponds to an ADAF model with $\dot{M}\sim 0.005\dot{M}_{\rm
Edd}$.  The dashed line is a thin disk model whose $\dot{M}$ has been
adjusted to fit the optical data.  This model fits poorly in the X-ray
band and is inconsistent with the EUV limit.  (From Narayan et
al. 1997a)}
\end{figure}

\begin{figure}
\includegraphics[width=3in,clip]{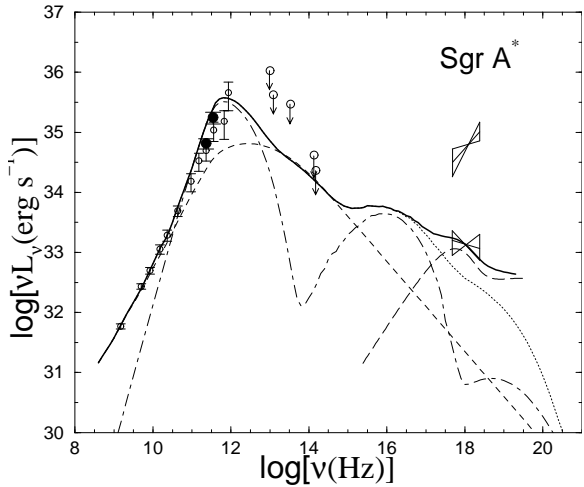}

\caption{The quiescent spectrum of Sgr A$^*$.  The radio data are from
Falcke et al. (1998; open circles) and Zhao et al. (2003; filled
circles), the IR data are from Serabyn et al. (1997) and Hornstein et
al. (2002), and the two ``bow-ties'' in the X-ray band correspond to
the quiescent (lower) and flaring (higher) data from Baganoff et
al. (2001, 2003).  The solid line is an ADAF model of Sgr A$^*$ in the
quiescent state.  The mass accretion rate is
$\dot{M}\sim10^{-6}\dot{M}_{\rm Edd}$ near the BH.  (From Yuan et al. 2003)}
\end{figure}

A typical accreting BH observed in the hard state has a luminosity on
the order of a few percent of Eddington.  At much lower luminosities,
we have the quiescent state, where the spectrum becomes noticeably
different.  Figures 2 and 3 show observations of a quiescent BHB, V404
Cyg, and a quiescent supermassive BH, the Galactic Center source Sgr
A$^*$.  Although these spectra look very different from the one shown
in Fig. 1, the ADAF model is able to fit these and other observations
of quiescent systems (Narayan, McClintock \& Yi 1996; Narayan et
al. 1997a; Yuan et al. 2003).  All it requires is a lower value of
$\dot{M}$, as appropriate for the lower luminosity.  The qualitative
features of the spectrum, e.g., a softening of the X-ray power-law
index (see Corbel, Tomsick \& Kaaret 2006), follow naturally.
However, a caveat is in order: In some cases, the X-ray emission in
quiescence may be from a jet lauched from the ADAF (\S3.6), rather
than from the ADAF itself (e.g., Yuan \& Cui 2005).

\subsection{Radiative Efficiency}

\begin{figure}
\includegraphics[width=3.3in,clip]{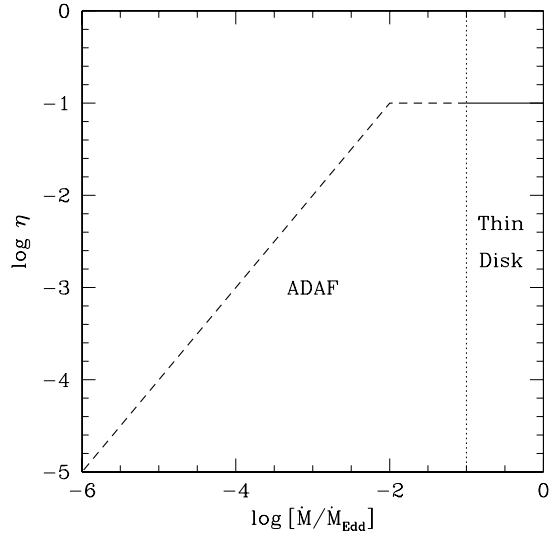}

\caption{Radiative efficiency $\eta$ of an accretion flow around a BH,
as a function of the Eddington-scaled mass accretion rate (Hopkins,
Narayan \& Hernquist 2006b).  The horizontal segment between 0.01 and
0.1 of the Eddington accretion rate shows a transition regime in which
a part of the accretion flow is an ADAF, but the radiative efficiency
is still large.  It might correspond to the Intermediate State, and
perhaps the upper end of the Hard State (\S3.5, Fig. 5).  Although
this plot is based on calculations shown in Fig. 11 of NY95b and
Fig. 13 of Esin et al. (1997), it is still very approximate.}
\end{figure}

As expressed in equation (\ref{ADAFdef}), the defining property of an
ADAF is that it is radiatively inefficient,
\begin{equation}
{\rm ADAF}: \qquad \eta\equiv L/\dot{M}c^2 \ll 0.1  .
\label{eta}
\end{equation}
Calculations show that the ADAF solution is possible only for low mass
accretion rates.  Specifically, only when $L\ \lta\ 0.1L_{\rm Edd}$ is
the gas density low enough to permit a two-temperature plasma (NY95b;
Narayan 1996; Esin et al. 1997).  Near the critical luminosity $L_{\rm
crit}$ or critical mass accretion rate $\dot{M}_{\rm crit}$ at which
the ADAF solution first becomes viable, the radiative efficiency is,
by continuity, not very different from that of a thin disk:
$\eta\sim0.1$.  However, with decreasing $\dot{M}$, the efficiency
decreases.  Very roughly, we estimate (Fig. 4)
\begin{equation}
\eta \sim 0.1\left(\dot{M}/0.01\dot{M}_{\rm Edd}\right), \quad
\dot{M} < 0.01\dot{M}_{\rm Edd}.
\end{equation}
(We note that for the different
prescription for $\delta$ used by Sharma et al. 2006, $\eta$ falls
rapidly only for $\dot{M} \ \lta\ 10^{-4} \dot{M}_{\rm Edd}$.)

The extreme inefficiency of an ADAF at very low accretion rates is
critical for understanding the peculiar properties of accreting BHs in
the quiescent state (as first discussed by Narayan et al.  1996 for
BHBs and Narayan et al. 1995 for supermassive BHs).  We do not discuss
this topic here, but point the reader to other reviews for details
(e.g., Quataert 2001; Narayan 2002, 2005).  The quiescent state has
also played a major role in our efforts to test for the presence of an
event horizon in BHBs.  This is a key topic of \S4.

\subsection{Spectral Regimes}

\begin{figure}
\includegraphics[width=3.7in,clip]{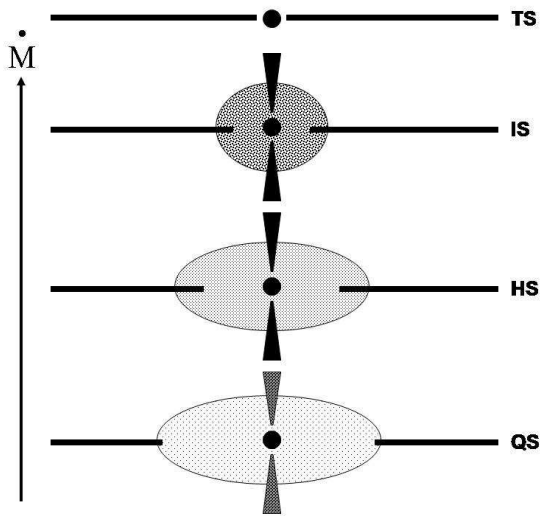}

\caption{Configuration of the accretion flow in different spectral
states, shown schematically as a function of the mass accretion rate
$\dot{M}$ (based on Esin et al. 1997).  The ADAF is represented by the
hatched ellipses, with the intensity of hatching indicating the
density of the hot gas.  The horizontal lines represent a standard
thin disk.  The lowest panel shows the quiescent state (QS), which
corresponds to a very low mass accretion rate (say $<10^{-3}
\dot{M}_{\rm Edd}$), a weak jet (\S3.6), a low radiative efficiency
(Fig. 4), and a large transition radius (Fig. 6).  The second panel
from the bottom shows the hard state (HS), where the mass accretion
rate is higher ($\sim 10^{-3}-10^{-1.5}\dot{M}_{\rm Edd}$) but still
below the critical rate $\dot{M}_{\rm crit}$, the jet is stronger, the
radiative efficiency is somewhat larger, and the transition radius is
smaller.  The second panel from the top shows the intermediate state
(IS), where $\dot{M}\sim 10^{-1.5}-10^{-1}\dot{M}_{\rm Edd}\ \lta\
\dot{M}_{\rm crit}$, the jet is even stronger, the radiative
efficiency is high $\sim0.1$, and the transition radius is fairly
close to the ISCO.  Finally, the top panel shows the thermal state
(TS), where there is no ADAF, the thin disk extends down to the ISCO,
$\eta\sim0.1$, and there is no jet.  Esin et al. (1997) included a
tentative proposal for the so-called very high state (see also Done et
al. 2007), now called the steep power-law state (SPL, McClintock \&
Remillard 2006), but we do not include this still-mysterious state
here.}
\end{figure}

Based on the properties of the ADAF solution discussed above, Narayan
(1996) proposed a simple model for understanding the spectral states
of accreting BHs.  This picture was developed in detail by Esin et
al. (1997).  The basic idea is illustrated in Fig. 5.

The key parameter that determines the spectral state of an accreting
BH is $\dot{M}$.  When $\dot{M}>\dot{M}_{\rm crit}$, only the thin
disk solution is available, and so the accretion flow is in the form
of a thin disk all the way down to the innermost stable circular orbit
(ISCO).  The system is then in the thermal state, and its spectrum is
well-described by the MCD model.

Once $\dot{M}$ falls below $\dot{M}_{\rm crit}$, both the thin disk
and ADAF solutions become viable (at least at small radii).  Now
accretion continues as a thin disk at radii larger than a transition
radius, $R > R_{\rm tr}$, but the flow switches to an ADAF at smaller
radii.  When the transition from a pure thin disk (thermal state) to a
disk-plus-ADAF configuration first occurs, i.e., when $\dot{M}$ is
just below $\dot{M}_{\rm crit}$, the ADAF is very small in size and we
have a more-or-less radiatively efficient flow, with $\eta\sim0.1$.
This corresponds to the so-called {\it intermediate state}.  Then, at
a somewhat lower $\dot{M}$, the ADAF expands a bit and $\eta$ is
modestly lower and we have the classic hard state.  Finally, when
$\dot{M}$ is much lower than $\dot{M}_{\rm crit}$, the ADAF becomes
much larger, $R_{\rm tr} \sim10^3-10^4R_S$, with $\eta\ll0.1$ (Narayan
et al. 1996, 1997a; Menou, Narayan \& Lasota 1999b).  This is the
quiescent state.

The above paradigm has proved durable (see Zdziarski \& Gierlinski
2004; Done et al. 2007).  In particular, considerable evidence has
accumulated that the thin disk retreats from the innermost stable
circular orbit (ISCO) to a large radius in the quiescent state.  The
evidence is strongest in transient BHBs and CVs.  The spectra of
quiescent BHBs show absolutely no sign of any soft blackbody-like
X-ray emission from a thin disk at small radii (Narayan et al. 1996,
1997a; McClintock, Narayan \& Rybicki 2004).  Timing properties also
indicate a large a large size for the ADAF (Hynes et al. 2003; Shahbaz
et al. 2005). In additional, theoretical arguments indicate that the
thermal-viscous disk instability, which causes the transient behavior
in these systems, is incompatible with observations unless the disk is
severely truncated in the quiescent state (Lasota, Narayan \& Yi
1996b; Hameury, Lasota \& Dubus 1999; Lasota 2001, 2008; Dubus,
Hameury \& Lasota 2001; Yungelson et al. 2006).  There are also
indications from the time delay between the optical and X-ray
outbursts in the BHB GRO J1655--40 (Orosz et al. 1997; Hameury et
al. 1997) that the cool disk is truncated at a large radius in the
quiescent state.

In the case of supermassive BHs in the quiescent state, there is no
feature in the spectrum that might be associated with a thin disk,
suggesting that there is no disk at all; examples are Sgr A$^*$
(Narayan et al. 1998a; Narayan 2002) and M87 (Di Matteo et al. 2000,
2003).  Intermediate luminosity AGN do exhibit optical emission from a
disk, but the ``big blue bump'' is much less pronounced than in
high-luminosity AGN (Ho 1999); this suggests that the disk is
truncated at a radius $\sim10-100R_S$ and the interior is filled with
an ADAF (Gammie, Narayan \& Blandford 1999; Quataert et al. 1999).
Incidentally, the thermal-viscous disk instability does not appear to
operate in AGN disks (Menou \& Quataert 2001; Hameury, Lasota \&
Viallet 2007).

In the more luminous hard state, again, there is considerable spectral
evidence that the disk is truncated at a transition radius outside the
ISCO, and that the inside is filled with an ADAF-like hot flow.  The
most spectacular example is the BHB XTE J1118+480, for which
observations carried out in the hard state had unprecedented spectral
coverage.  The observations are fit well with an ADAF model, with a
transition radius at $\sim 50 R_S$ (Esin et al. 2001).  The model even
fits the complicated timing behavior of the source (Yuan, Cui \&
Narayan 2005).  It is hard to imagine that the same data could be
explained with any model in which a cool disk extends down to the
ISCO.  Done et al. (2007) review spectral observations of a number of
other BHBs where again the data require a truncated disk.  Nemmen et
al. (2006) show that both the spectrum and the double-peaked Balmer
line profile of the LINER source NGC 1097 are consistent with a disk
truncated at a few hundred $R_S$.

In an interesting study of Cyg X--1, Gilfanov, Churazov \& Revnivtsev
(1999; see Cui et al. 1999; Done et al. 2007; for discussions of other
sources) found that, as the characteristic frequency in the
variability spectrum of the source increases, the power-law tail in
the spectral energy distribution steepens and the amplitude of the
reflection component in the spectrum increases.  This is exactly what
one expects when the transition radius between the outer cool disk and
the inner hot ADAF varies.  With decreasing transition radius, (i) the
noise frequency (which is likely related in some fashion to the
Keplerian frequency at the transition radius) should increase, (ii)
the hot medium should be cooled more effectively by soft photons from
the disk, giving a steeper power-law tail, and (iii) the solid angle
subtended by the cool disk at the ADAF should increase, and there
should be a larger reflection component.  Zdziarski, Lubinski \& Smith
(1999) have noted that a correlation between spectral slope and
reflection is commonly seen in both BHBs and AGN.

A few BHBs in the hard state have been found to show a soft
blackbody-like component in their spectra (Balucinska-Church et
al. 1995; Di Salvo et al. 2001; Miller et al. 2006a,b; Ramadevi \&
Seetha 2007; Rykoff et al. 2007).  This could be interpreted as
evidence that the thin disk extends all the way down to the
ISCO\footnote{However, Di Salvo et al. (2001) estimate from their
observations of Cyg X--1 that the disk inner edge is located at tens
of $R_S$, not at the ISCO}.  Some authors have noted that it is
possible for a thin disk to evaporate to an ADAF at a relatively large
radius and for the hot gas to then re-condense into a thin disk at
small radii (Rozanska \& Czerny 2000; Liu et al. 2007; Mayer \&
Pringle 2007)\footnote{Models of this kind may give hard spectra
without a significant decrease in radiative efficiency.  This would be
consistent with the transition region between mass accretion rates of
0.01 and 0.1 of Eddington shown in Fig. 4, where we have suggested an
ADAF with high radiative efficiency may be present.}.  Such models
might explain the occurrence of a soft spectral component in some hard
state sources.  However, we note that the soft component typically has
only 10\% of the total observed luminosity.  It is hard to understand
how a radiatively efficient thin disk located at the ISCO could be
such a minor contributor to the emitted radiation.

This difficulty is highlighted in the work of D'Angelo et al. (2008).
They point out that, in any model of the hard state, there will be
considerable interaction between the hot gas which produces the hard
X-rays and cool gas that may be present in a conventional disk.  The
interaction will occur via X-ray irradiation (unless one has large
outward beaming, which seems unlikely, e.g., Narayan \& McClintock
2005) as well as particle bombardment.  Using a prototype model for
ion bombardment (Deufel et al. 2002; Spruit \& Deufel 2002; Dullemond
\& Spruit 2005), D'Angelo et al. (2008) show that a weak soft
component in the X-ray spectrum arises quite naturally when the thin
disk is truncated at $\sim15-20 R_S$\footnote{This is a reasonable
location for the transition radius in the hard state (Esin et
al. 1998; Done et al. 2007)}.  The model fits the observations
surprisingly well; in fact, the same model would predict a much
stronger soft component, and would strongly disagree with the
observations, if the cool disk were to extend down to the ISCO.

\begin{figure}
\includegraphics[width=3.1in,clip]{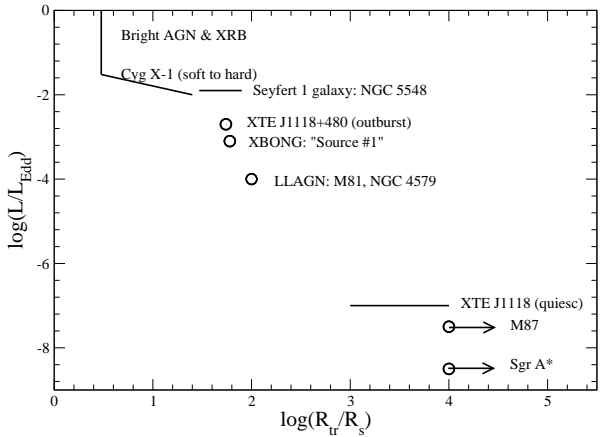}

\caption{Estimated location of the dimensionless transition radius
$R_{\rm tr}/R_S$, plotted as a function of the Eddington-scaled
accretion luminosity $L/L_{\rm Edd}$, for a sample of BHBs and
low-luminosity AGN (Yuan \& Narayan 2004).  The individual estimates
of $R_{\rm tr}$ are obtained by fitting spectral observations and are
very uncertain.  Nevertheless, there seems to be a trend of increasing
transition radius with decreasing luminosity.}
\end{figure}

One of the most vexing problems in the theory of ADAFs is that we do
not have a robust method of estimating the location of the transition
radius $R_{\rm tr}$.  From the earliest studies (e.g., NY95b; Meyer \&
Meyer-Hofmeister 1994), it has been plausibly argued that $R_{\rm tr}$
increases with decreasing $\dot{M}$.  However, reliable predictions of
the exact dependence have proved difficult, although a number of
studies have come up with semi-quantitative results (Meyer \&
Meyer-Hofmeister 1994; Honma 1996; Rozanska \& Czerny 2000; Meyer, Liu
\& Meyer-Hofmeister 2000; Spruit \& Deufel 2002; Mayer \& Pringle
2007).  Yuan \& Narayan (2004) tried to use observations to deduce the
run of $R_{\rm tr}$ with accretion luminosity.  Figure 6 shows their
results.

Another issue that has been discussed recently is hysteresis in the
transition between the thermal state and the hard state.
Specifically, with increasing $\dot{M}$, the hard state survives up to
fairly high luminosities $\sim0.1-0.2L_{\rm Edd}$, whereas with
decreasing $\dot{M}$, the transition from the thermal state to the
hard state occurs at a much lower luminosity $\sim 0.02-0.05L_{\rm
Edd}$ (Miyamoto et al. 1995; Maccarone \& Coppi 2003; Zdziarski et
al. 2004; Done et al. 2007).  Meyer-Hofmeister, Liu \& Meyer (2005)
have provided a plausible explanation.  Their disk evaporation model,
coupled with Compton-cooling of the hot electrons in the ADAF,
naturally produces a hysteresis in the location of the transition
radius.

The brightest systems in the hard state ($L\ \gta\ 0.1L_{\rm Edd}$)
are difficult to model with the standard ADAF model.  A variant of the
ADAF solution --- a natural extension of the model --- called the
luminous hot flow (LHAF; Yuan 2001, 2003; Yuan \& Zdziarski 2004; Yuan
et al. 2007; see also Machida, Nakamura \& Matsumoto 2006), looks
promising for modeling these sources.

\subsection{Outflows and Jets}

NY94, NY95a discovered an unexpected property of the ADAF solution:
{\it the accreting gas in an ADAF has a positive Bernoulli
parameter}\footnote{Technically, it should be called the Bernoulli
``function'' since the quantity varies with radius, but we use
Bernoulli ``parameter'' since that was the term used in NY94}, which
is defined as the sum of the kinetic energy, potential energy and
enthalpy,
\begin{equation}
Be = v^2/2 -GM/R +w.  
\end{equation} 
A positive Bernoulli constant means that the gas is not bound to the
BH.  (This is not surprising, since the gas is not losing energy
through radiation.)  Therefore, the above authors suggested that ADAFs
should be associated with strong winds and jets\footnote{There has
been a tendency in the literature to ignore these early papers and to
credit Blandford \& Begelman (1999) for establishing the connection
between ADAFs and outflows.  We note that the abstract of the first
paper on ADAFs, NY94, states: ``Further, the Bernoulli parameter is
positive, implying that advection-dominated flows are susceptible to
producing outflows...We suggest that advection-dominated accretion may
provide an explanation for...the widespread occurrence of outflows and
jets in accreting systems.''  In the second paper, NY95a, the title
itself mentions outflows, ``Advection-Dominated Accretion:
Self-Similarity and Bipolar Outflows,'' and an entire paragraph of the
abstract is devoted to the topic.} (see also Meier 2001).

Strong outflows have been seen in numerical simulations of ADAFs.  The
first indications came from 2D and 3D hydrodynamic simulations (Stone,
Pringle \& Begelman 1999; Igumenshchev \& Abramowicz 2000;
Igumenshchev, Abramowicz \& Narayan 2000), but it was soon confirmed
in MHD simulations as well (Stone \& Pringle 2001; Hawley \& Balbus
2002; Igumenshchev, Narayan \& Abramowicz 2003; Pen, Matzner \& Wong
2003; Machida, Nakamura \& Matsumoto 2004; McKinney \& Gammie 2004;
Igumenshchev 2004).

Apart from producing a gas-dominated, large-scale outflow, MHD
simulations of ADAFs also have a second distinct outflow component
along the axis in the form of a collimated, Poynting-dominated,
relativistic jet (McKinney 2005, 2006).  The original suggestion of
NY94 that ADAFs would have outflows and jets has thus been confirmed
by these simulations.  Nevertheless, some authors have disputed a
connection between a positive Bernoulli parameter and an outflow since
it is possible to come up with explicit models that have a positive
Bernoulli parameter but no outflow (Paczy\'nski 1998; Abramowicz et
al. 2000).

Observational evidence for the association of nonthermal relativistic
jets with ADAFs has accumulated in recent years with the discovery of
radio emission in virtually every BHB in the hard state (Corbel et
al. 2000; Fender 2001; Fender, Belloni \& Gallo 2004; Fender \&
Belloni 2004).  The radio emission is generally too bright to be
produced by thermal electrons in the accretion flow.  It is therefore
very likely to come from nonthermal electrons in a jet; in fact, radio
VLBI imaging has revealed a resolved jet in a few sources.  Thus, it
is now observationally well-established that the hard state/ADAF is
associated with relativistic jets.  According to the discussion in
\S3.5, the quiescent state also has an ADAF and should have a
(weaker) jet.  Indeed, radio emission has been seen from two quiescent
systems, V404 Cyg (Hjellming et al. 2000; Gallo, Fender \& Hynes 2005)
and A0620--00 (Gallo et al. 2006), confirming this expectation.

How much of the X-ray emission in an ADAF system comes from the
accretion flow and how much from the jet?  A strong correlation has
been seen between radio and X-ray luminosity in the hard state and
quiescent state (Corbel et al. 2003; Gallo, Fender \& Pooley 2003).
At first sight this might suggest that the X-ray must also come from
the jet (e.g., Falcke, K\"ording \& Markoff 2004).  However, since the
jet flows out of the ADAF and is thus highly coupled to it, any model
in which the X-ray emission is from the ADAF and radio is from the jet
is equally compatible with the observations.  In recent times, several
authors have come down in favor of an ADAF origin for the X-rays
(e.g., Heinz \& Sunyaev 2003; Merloni, Heinz \& Di Matteo 2003; Heinz
2004; Heinz et al. 2005; Yuan et al. 2005).

Additional evidence on this issue comes from hard state spectra
indicating thermal emission (e.g., see Fig. 1), which is very
different from the nonthermal power-law emission one expects from a
jet.  Markoff, Falcke \& Fender (2001), among others, have argued that
the X-ray emission is due to synchrotron emission from a carefully
tuned distrubution of nonthermal electrons.  However, Zdziarski et
al. (2003) and Zdziarski \& Gierlinski (2004) showed that the
high-energy cutoff in the spectra of BHBs in the hard state, which
arises naturally in a thermal ADAF-like model, is very difficult to
reproduce in a nonthermal synchrotron model.  A variant of the jet
model assigns the X-ray emission to thermal electrons in the ``base of
the jet'' (Markoff, Nowak \& Wilm 2005), but the discussion then
becomes semantic.  The base of the jet is surely embedded in the
underlying ADAF and it is not clear that one gains anything by simply
relabeling the radiation from the ADAF as jet emission.  The issue is
discussed in greater depth in Narayan (2005), who advocates a
`jet-ADAF' model (as described in Yuan et al. 2005; Malzac, Merloni \&
Fabian 2004), in which the high energy X-ray emission in the hard
state comes from the ADAF and the low-energy radio (and infrared)
emission comes from a jet.

The situation is less clear-cut in the quiescent state, where
nonthermal emission from the jet might dominate even in X-rays (Yuan
\& Cui 2005; Wu, Yuan \& Cao 2007).  It is interesting, however, that
the most quiescent system we know, Sgr A$^*$, shows no sign of a jet.
Radio images with a resolution of $\sim15R_S$ appear to be jet-free
(Shen et al. 2005), while the quiescent X-ray emission is spatially
resolved and appears to be from thermal gas near the Bondi radius at
$\sim10^5R_S$, not from a jet (Baganoff et al. 2001, 2003).

Turning our attention now to the extended wind that is predicted from
an ADAF, and seen in numerical simulations, one consequence of the
wind is that, at each radius, a fraction of the accreting gas is lost
from the system.  Thus, the accretion rate itself varies with radius.
This is usually parameterized with an index $s$ such that $\dot{M}(r)
\propto r^{s}$.  Operationally, this means that, in the self-similar
regime, the gas density varies with radius as $\rho\propto r^{-3/2+s}$
rather than as $r^{-3/2}$ in the NY94 model.  Unfortunately, apart
from the rather general constraint that $s$ should lie between 0 (no
mass loss) and 1 (the limit of a CDAF, \S3.7), there is no good
theoretical estimate of the value of $s$.  We have to resort to
numerical simulations or observations.  The former tends to give
somewhat larger values, e.g., $s\sim0.7$ (Pen et al. 2003), while the
one case where the latter approach has been tried, Sgr A$^*$, yields a
smaller value, $s\sim0.3$ (Yuan et al. 2003).

\subsection{Convection}

Another general property of ADAFs was highlighted by NY94 (see also
Begelman \& Meier 1982): ADAFs have entropy increasing inward and
should be convectively unstable by the Schwarzschild criterion.
Numerical hydrodynamic simulations of ADAFs confirmed the presence of
strong convection (Igumenshchev et al. 2000; Narayan, Igumenshchev \&
Abramowicz 2000) and led to the development of an analytical
self-similar model called the convection-dominated accretion flow
(CDAF; Narayan et al.  2000; Quataert \& Gruzinov 2000).

The CDAF model employs a simplified one-dimensional treatment of the
accretion flow, in which all fluxes are assumed to be in the radial
direction.  The resulting density varies as $\rho \propto r^{-1/2}$.
In the language of \S3.6, a CDAF corresponds to $s=1$.  However,
technically, an idealized CDAF has no mass outflow, only an outward
flow of energy by convection; the energy is assumed to flow out into a
surrounding medium.  In practice, the flow is never perfectly
one-dimensional; 2D effects intrude and one expects substantial winds
and mass loss as well.  A real ADAF thus involves a complicated
interplay between convection and winds.

Magnetic fields introduce further complexity, and there has been some
discussion in the literature on whether MHD ADAFs do or do not have
real convection (Machida, Matsumoto \& Mineshige 2001; Balbus \&
Hawley 2002; Narayan et al. 2002b; Pen et al. 2003; Igumenshchev
2004).  The question has no practical significance, however, since
there is general agreement that ADAFs have both unstable entropy
gradients and strong magnetic stresses.

In the particular case of a spherical, non-rotating, MHD flow (the
magnetic analog of the Bondi problem), some complications are absent
and the problem becomes relatively clean.  Here one finds that
convection and magnetic fields strongly influence the accretion
physics, and the resulting flow is very different from the standard
Bondi solution (Igumenshchev \& Narayan 2002; Igumenshchev 2004,
2006).  This result is likely to have significant implications for
astrophysics; for instance, isolated neutron stars accreting from the
interstellar medium will be very much dimmer than one might expect
based on the Bondi solution (Perna et al. 2003).

\subsection{ADAFs Around Supermassive Black Holes}

The ADAF solution is essentially independent of the mass $M$ of the
central BH.  That is, if length and time are scaled by $M$ and the
accretion rate is scaled by the Eddington rate (also proportional to
$M$), then the same solution is valid for any BH mass.  Therefore, any
successful application of the ADAF model to a BHB system has immediate
consequences for supermassive BHs (SMBHs) in an equivalent state, and
vice versa.  This close connection between ADAFs in BHBs and ADAFs in
AGN was highlighted in Narayan (1996); it is also empirically obvious
from Fig. 6, which combines observations and models of BHBs and AGN.

The first SMBH to be modeled as an ADAF was the Galactic Center source
Sgr A$^*$ (Narayan et al. 1995).  This was soon followed by Fabian \&
Rees (1995), who suggested that quiescent nuclei in nearby giant
ellipticals (e.g., M87) must be accreting via ADAFs, and by Lasota et
al. (1996a), who argued that Low Ionization Nuclear Emission-line
Region sources (LINERs, e.g., NGC 4258) and low-luminosity active
galactic nuclei (LLAGN) must have ADAFs.  Both suggestions have turned
out to be correct (e.g., Reynolds et al. 1996; Mahadevan 1997; Gammie
et al. 1999; Quataert et al. 1999; Di Matteo et al. 2000, 2003;
Loewenstein et al. 2001; Ulvestad \& Ho 2001; Nemmen et al. 2006).  

Other authors have suggested that all of the following systems have
ADAFs: FRI sources (Baum, Zirbel \& O'Dea 1995; Reynolds et al. 1996;
Begelman \& Celotti 2004), BL Lac sources (Maraschi \& Tavecchio
2003), X-ray Bright Optically Normal Galaxies (XBONGs; Yuan \& Narayan
2004), and even some Seyferts (Chiang \& Blaes 2003).  All of these
sources are relatively low-luminosity AGN, where an ADAF is likely to
be present (Figs. 5, 6).

As discussed in \S3.6, a feature of the ADAF model is the presence of
outflows and jets.  This implies that ADAF systems should generally be
radio-loud.  This is certainly the case for the FRI and BL Lac sources
mentioned above.  Ho (2002; see also Nagar et al. 2000; Falcke et
al. 2000) presents in Fig. 5b of his paper a very interesting plot of
radio loudness, defined as the ratio of the flux at 6 cm to the flux
in the optical B band versus Eddington-scaled luminosity, for a
sample of galactic nuclei.  Virtually every source in the plot with an
Eddington-ratio below about 0.01 is radio loud, which is perfectly
consistent with our expectation that all such sources should have
ADAFs.  Moreover, the degree of radio loudness increases with
decreasing Eddington ratio, again consistent with our expectation that
the ADAF should become more and more dominant with decreasing
$\dot{M}$.  The majority of sources in Ho's plot that have $L/L_{\rm
Edd} > 0.1$ are radio-quiet, as we would expect from Fig. 5 if these
systems have cool disks.  However, a small minority of these bright
sources do show powerful jets (they are generally FRII sources).  The
exact nature of the accretion in these sources is unclear.  A more
complete plot, with updated data, can be found in Sikora, Stawarz \&
Lasota (2007).

The jet and the extended wind from an ADAF carry with them substantial
kinetic energy.  This energy will be dumped into the external medium
and will have important consequences.  In the context of galaxy
formation, there has been discussion recently of the so-called ``radio
mode'' of accretion (Croton et al. 2006) in which outflowing energy
from an accreting SMBH in the galactic nucleus heats up the
surrounding medium.  This kind of AGN feedback can lead to various
effects such as shutting off accretion and/or star formation (Di
Matteo et al. 2005; Hopkins et al. 2006a).

It is worth noting that this radio mode is nothing other than the ADAF
mode of accretion reviewed in this article.  Investigators of AGN
feedback may find it profitable to study the considerable work that
has been done on ADAFs over the last fifteen years.

\section{The Black Hole Event Horizon}

The first BH, Cygnus X-1, was identified and established in 1972 via a
measurement of its mass, which was shown to be too large for a neutron
star (NS).  The surest evidence for the existence of BHs continues to
be through dynamical mass measurements.  We now know of 20 additional
compact binary X-ray sources (McClintock \& Remillard 2006; Orosz et
al. 2007) with primaries that are too massive to be a NS or any stable
assembly of cold degenerate matter, assuming that GR is valid.
Similarly, dynamical data have established the existence of
supermassive BHs, most notably in the nucleus of our Milky Way Galacy
(Sch\"odel et al. 2002; Ghez et al. 2005a) and in NGC 4258 (Miyoshi et
al. 1995).

Are these compact objects genuine BHs -- pockets of fully collapsed
matter that are walled off from sight by self gravity and that, like a
shadow, reveal no detail --- or are they exotic objects that have no
event horizons but manage to masquerade as BHs?  Most astrophysicists
believe that they are genuine BHs.  There are several reasons for this
confidence, the most important being that BHs are an almost inevitable
prediction of GR.  However, this argument is circular because it
presumes that GR is the correct theory of gravity.  Furthermore, it
ignores the many exotic alternatives to BHs that have been suggested.
Thus, the collapsed objects that we refer to throughout as ``black
holes'' are strictly speaking dynamical BH candidates.  The current
evidence for BHs is not decisive, nor can dynamical measurements be
expected to make it so.  We now consider some approaches aimed at
establishing that these dynamical BHs are genuine.

The defining property of a BH is its event horizon.  Demonstrating the
existence of this immaterial surface would be the certain way to prove
the reality of BHs.  Unfortunately, unlike any ordinary astronomical
body such as a planet or a star of any kind, it is quite impossible to
detect radiation from the event horizon's surface of infinite
redshift.  (Hawking radiation is negligibly weak for massive
astrophysical BHs.)  Nevertheless, despite the complete absence of any
emitted radiation, it is possible to marshal strong {\it
circumstantial evidence} for the reality of the event horizon.  The
fruitful approaches described below are based on comparing X-ray
binary systems that contain BH primaries with very similar systems
that contain NS primaries.  Such investigations are motivated by the
simple fact that the termination of an accretion flow at the hard
surface of a NS has observational consequences.

The earliest and strongest evidence for the event horizon is based on
the faintness in quiescence of BH transient systems relative to
comparable NS systems.  In \S4.1, we discuss the physical arguments
that underpin this evidence, and in \S4.2 we present the comparative
luminosity data for BHs and NSs.  Alternative explanations for the
lower luminosities of the BH systems, which do not involve the event
horizon, are considered in \S4.3.  In \S4.4, we present three
independent and additional arguments for the existence of the event
horizon, which are again rooted in comparing BH and NS X-ray binaries.
In \S4.5, we discuss the extreme faintness and properties of Sgr A*
and the evidence that this supermassive BH has an event horizon, and
in \S4.6 we argue that even very exotic objects with enormously strong
surface gravity (e.g., gravastars) cannot escape our arguments for the
event horizon.

\subsection{Accretion and the Event Horizon}

A test particle in a circular orbit at radius $R$ around a mass $M$
has, in the Newtonian limit, kinetic energy per unit mass equal to
$GM/2R$, and potential energy equal to $-GM/R$.  In the context of a
gaseous accretion disk, this means that a gas blob at radius $R$
retains $50\%$ of its potential energy as kinetic energy.  The
remaining 50\% was transformed into thermal energy during the viscous
accretion of the blob to its current radius (Frank et al. 2002).

We now turn to consider accretion on to the material surface of a NS.
If the mass $M$ at the center of an accretion disk has a surface, then
the kinetic energy in the accreting gas will be converted to thermal
energy in a viscous boundary layer and radiated (Frank et al. 2002).
In addition, the residual thermal energy that the gas possesses when
it reaches the inner edge of the disk will also be radiated from the
surface.

If accretion occurs via an ADAF, the luminosity from the accretion
disk $L_{\rm acc}$ and that from the stellar surface $L_{\rm surf}$
will satisfy
\begin{eqnarray}
L_{\rm acc} &\ll& GM\dot{M}/2R_*, \\
L_{\rm surf} &\approx& GM\dot{M}/R_* \gg L_{\rm acc},
\label{ADAFsurf}
\end{eqnarray}
where $R_*$ is the radius of the stellar surface.  The accretion
luminosity is small because the flow is radiatively inefficient.
Therefore, essentially all the potential energy of the accreting gas
remains in the gas in the form of thermal and kinetic energy.  If the
central object has a surface, e.g., it is a NS, the total luminosity
we observe will be equal to $L_{\rm surf}$.  However, if the object
has an event horizon, i.e., it is a BH, there will be no radiation
from a stellar surface and the luminosity we observe will only be
equal to $L_{\rm acc}$.  Thus, we expect
\begin{equation}
{\rm ADAF}: \quad L_{\rm NS} \gg L_{\rm BH}.
\label{ADAFlum}
\end{equation}

\begin{figure}
\includegraphics[width=3.3in,clip]{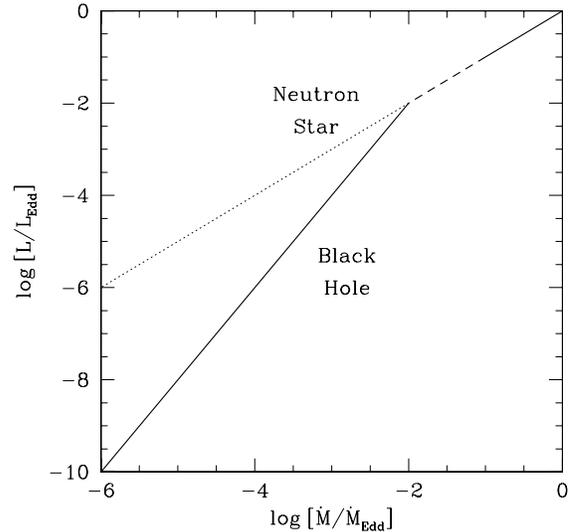}

\caption{Predicted luminosities of accreting BHs and NSs as a function
of mass accretion rate.  This plot is based on the accretion
efficiency prescription shown in Fig. 4.  Because a NS has a surface,
the total luminosity is always $\sim GM_{\rm NS}\dot{M}/R_{\rm NS}
\sim 0.1\dot{M}c^2$, regardless of whether the accretion flow is
radiatively efficient or not.  In the case of a BH, however, the
radiative efficiency $\eta$ of the accretion flow is of paramount
importance, and the observed luminosity becomes $\ll0.1\dot{M}c^2$
when the accretion rate is highly sub-Eddington.}
\end{figure}

Figure 7 shows schematically the luminosity difference we predict
between a NS and a BH.  This plot is based on the accretion efficiency
estimate shown in Fig. 4.  Especially at low mass accretion rates, say
$\dot{M}\ \lta\ 10^{-4}\dot{M}_{\rm Edd}$ (deep quiescent state), we
expect a huge luminosity difference between NSs and BHs.  Efforts to
test this prediction are described in \S4.2.

If the accretion disk is radiatively efficient, then the gas has
little thermal energy when it reaches the inner edge of the disk.  In
this case, we expect\footnote{The result given here is for a
non-spinning accretor.  If the central mass is spinning, the boundary
layer luminosity from the surface will be somewhat smaller.  However,
it requires very fine tuning to make $L_{\rm surf} \ll L_{\rm acc}$.
Moreover, if the central mass spins too close to the ``break-up''
limit, the extra luminosity due to the surface can actually exceed
$GM\dot{M}/2R_*$ by a large factor (Popham \& Narayan 1991).}
\begin{eqnarray}
L_{\rm acc} &\approx& GM\dot{M}/2R_*, \\
L_{\rm surf} &\approx& GM\dot{M}/2R_* \sim L_{\rm acc}.
\label{surf2}
\end{eqnarray}
The above result is based on a Newtonian analysis and is okay so long
as the central object has a radius $R_*\ \gta\ R_{\rm ISCO}$, the
radius of the ISCO.  For more compact objects, we must allow for the
fact that, inside the ISCO, the accreting gas no longer spirals in by
viscosity but free-falls in the gravitational potential of the central
mass.  Now we have
\begin{eqnarray}
L_{\rm acc} &\approx& 0.1\dot{M}c^2, \\
L_{\rm surf} &\approx& GM\dot{M}/R_*- L_{\rm acc} > L_{\rm acc}.
\label{surf3}
\end{eqnarray}
The accretion luminosity is limited by the binding energy of the gas
at the ISCO, which gives a radiative efficiency $\eta\sim0.1$.  On the
other hand, the total energy budget is $GM\dot{M}/R_*$, which means
that a larger fraction of the luminosity is released at the surface.
(The free-falling gas inside the ISCO, crashes on the surface and
releases its energy in a shock.)  In the limit when $R_*$ is
arbitrarily close to $R_S$ (cf., the discussion of gravastars in
\S4.6), the total luminosity is equal to $\dot{M}c^2$, i.e., the
entire rest mass energy of the accreting gas is converted to radiation
(100\% radiative efficiency).  Note that all luminosities discussed
here refer to measurements by an observer at infinity.

Combining the results in equations (\ref{ADAFsurf}), (\ref{surf2}) and
(\ref{surf3}), we see that, regardless of the radius of the accretor
and whether the accretion flow is radiatively efficient or
inefficient, we expect
\begin{equation}
{\rm Any~Accretion~Flow:} \qquad L_{\rm surf} \ \gta\ L_{\rm acc}.
\label{Lsurfacc}
\end{equation}
Although this relation is weaker than (\ref{ADAFsurf}), it can still
be used in favorable cases to test for the presence of an event
horizon (\S4.5).

\subsection{Evidence for the Event Horizon: Luminosities of Quiescent X-ray
  Binaries}

BH and NS transient binary systems are very similar in many respects,
and it is reasonable to expect that their mass accretion rates and
luminosities would be comparable under similar conditions.  There is,
however, one important qualitative difference between the two kinds of
object --- NSs have surfaces whereas BHs do not.  Often, this
difference is not important.  However, as discussed in \S4.1 (see
eq. \ref{ADAFlum}), when accretion occurs via an ADAF, a NS binary
should be much more luminous than a BH binary (NY95b).  The difference
will be especially large in the quiescent state, when the accretion
flow is radiatively extremely inefficient (Fig. 7).

Narayan, Garcia \& McClintock (1997b) and Garcia, McClintock \&
Narayan (1998) collected available X-ray data on quiescent NS and BH
transients and showed that BH systems are consistently fainter than NS
systems.  This was the first indication that the objects that
astronomers call ``black holes'' are indeed genuine BHs with
event horizons.

\begin{figure}
\includegraphics[width=3.3in,clip]{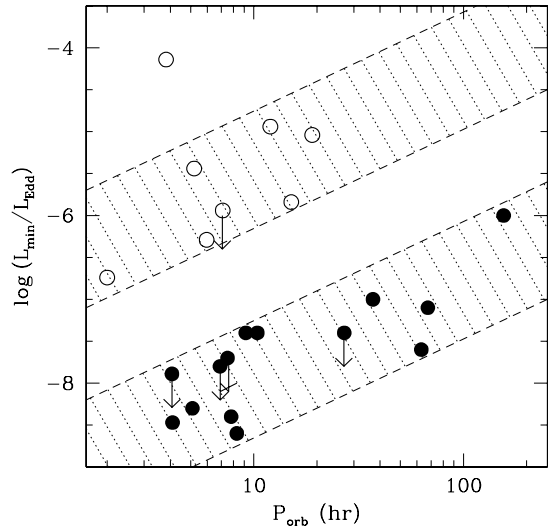}
\caption{Eddington-scaled luminosities (0.5--10 keV) of BH transients
({\it filled circles}) and NS transients ({\it open circles}) versus
the orbital period (see Garcia et al. 2001; McClintock et al. 2004).
Only the lowest quiescent detections or {\it Chandra/XMM} upper limits
are shown.  The plot shows all systems with known orbital periods,
which have optical counterparts and good distance estimates.  The
diagonal hatched areas delineate the regions occupied by the two
classes of sources and indicate the observed dependence of luminosity
on orbital period.  Note that the BH systems are on average nearly 3
orders of magnitude fainter than the NS systems with similar orbital
periods. }
\end{figure}

\begin{figure}
\includegraphics[width=3.3in,clip]{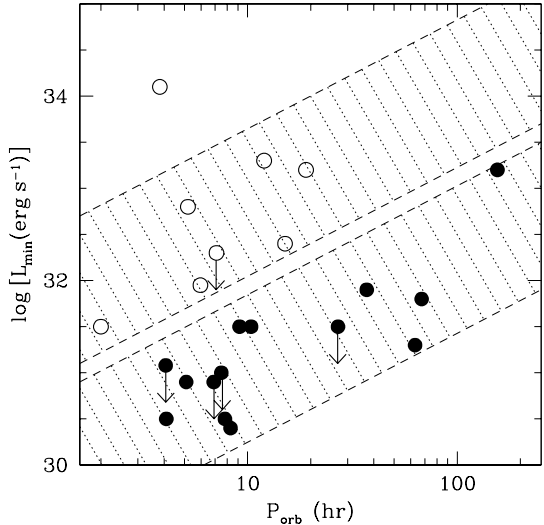}
\caption{Same as Fig.\ 8 except that shown here are the observed
luminosities without Eddington scaling.  In this representation, the
BH systems are on average a factor of $\sim 100$ fainter than NS
systems with similar orbital periods. }
\end{figure}

Lasota \& Hameury (1998) made the important point that it is necessary
to compare quiescent NS and BH transients at similar mass accretion
rates, and the surest way to achieve this is to plot luminosities as a
function of the binary orbital period $P_{\rm orb}$ (the relevant
arguments are outlined in Menou et al. 1999b; Lasota 2000, 2008).
Since 1999, this has been the standard way of plotting the data (Menou
et al. 1999a; Garcia et al. 2001; Narayan, Garcia \& McClintock 2002a;
Hameury et al. 2003; McClintock et al. 2004).  Figure 8 shows the
current status of the comparison, with the luminosities of quiescent
BH and NS transients plotted in (most-appropriate) Eddington units.
Fig. 9 shows the same data without the luminosities being scaled.  It
is clear that, for comparable orbital periods, the BH systems are 2 to
3 orders of magnitude fainter than their NS cousins.

As discussed in \S4.2, a radiatively inefficient ADAF provides a
natural explanation for the large luminosity deficit of the BH
systems.  In this model, the gas approaches the center with a large
amount of thermal energy.  A BH is dim because the bulk of this
thermal energy is trapped in the advective flow, passes through the
event horizon, and is lost from sight.  On the other hand, a NS is
bright because the thermal energy is radiated from its surface.

\subsection{Other Explanations for the Relative Faintness of Quiescent
  Black Hole Binaries}

A number of alternative models have been put forward in an attempt to
rationalize the luminosity differences between quiescent NS and BH
systems.  Notice below that some of these models generate X-ray
emission without invoking accretion at all.  We first consider a
recent challenge centered on the discovery of the extremely
low-luminosity NS transient 1H 1905+00.  Next, we discuss the
possibility that the bulk of the accretion power is channeled into a
steady jet rather than into X-ray emission.  We then consider several
diverse models that are discussed in further detail in Narayan et
al. (2002a).

{\it The case of 1H 1905+00:} Jonker et al.\ (2007) claim that the
extremely low X-ray luminosity of the NS transient and type I burst
source 1H 1905+00 (hereafter, H1905) undermines the evidence
summarized in \S4.2 for the existence of the event horizon.  The
luminosity of this NS ($\approx 2 \times 10^{30} ~{\rm erg\,s^{-1}}$;
$D = 10$~kpc), is the same as the luminosity of A0620--00 and several
other short-period ($P_{\rm orb} \approx 4-12$~hr) BH systems.  Based
on this result, Jonker et al.\ assert that the evidence for event
horizons is ``unproven.''  However, they ignore a key point of our
argument: Namely, as discussed above and illustrated in Figures 8 and
9, the case for event horizons depends critically on comparing BH and
NS systems with {\it similar orbital periods}.

We first note that the orbital period of H1905 is unknown.  More
importantly, this unknown period is believed to be very short -- so
short that there are no BH systems with comparable periods.  This
conclusion is based on deep optical imaging data and the lack of a
counterpart.  Jonker et al.\ (2007) conclude that the secondary ``can
only be a brown or a white dwarf,'' and that the system is probably an
ultracompact binary.  Hence the orbital period is likely to be tens of
minutes, far less than the shortest BH binary period of 4.1~hr.  

Thus, there is no BH system comparable to H1905.  And there is no way
to usefully predict the mass accretion rate, as Jonker et al. conclude
in the paper's final sentence.  The rough trend of luminosity with
period indicated in Figure 9 for NSs might imply a very low luminosity
for H1905, as observed, but such an extrapolation is unwarranted.  See
Lasota (2007, 2008) for a more detailed discussion of this and other
issues.

Jonker et al.\ (2007) additionally argue that an unknown amount of
mass transfered from the secondary might be lost from the system in
winds or jets.  Their one example of a wind, viz., an extraordinary
mass-loss episode observed in GRO J1655-40 (Miller et al. 2006c) when
the source was in outburst and had a luminosity $\sim 10^6$ times the
quiescent level, is quite inappropriate.  Also, their mention of
possible mass expulsion for quiescent NSs via the propeller mechanism
(Lasota \& Hameury 1998; Menou et al. 1999a), if effective, would
reduce the luminosities of NSs and would only strengthen our argument.

{\it Jet outflows:} Fender et al. (2003) argue that transient BHs at
low accretion rates ($L/L_{\rm Edd} < 10^{-4}$) ``should enter
`jet-dominated' states,'' in which the majority of the accretion power
drives a radiatively-inefficient jet.  As we have seen in \S3.6, there
is good evidence for this.  For instance, the presence of a radio jet
has been reasonably well established in quiescence for A0620--00
(Gallo et al. 2006), the closest and one of the least luminous of the
BHs in question.  Fender et al. (2003) appeal to the empirical result
that BHs are $\sim 30$ times as `radio loud' as NSs at similar
accretion rates to deduce that quiescent BHs should be $\sim 100$
times {\it less} luminous in X-rays than quiescent NSs.  Their
argument is a little convoluted since, if the radio and X-ray emission
in quiescence come from the jet (see \S3.6), one would think the BHs
would be $\sim 30$ times {\it more} luminous, not 100 times less
luminous.  Therefore, naively, the jet argument only makes the
discrepancy in Figs. 8 and 9 more severe.  Fender et al. (2003) get
around this difficulty by postulating, in addition to different jet
efficiencies, also different origins for the X-rays seen in quiescent
BHs and NSs.

It is unsatisfying that the reason for the difference in jet activity
between BHs and NSs ``remains unclear.''  At bottom, the authors are
suggesting that in quiescence the BHs are in the `jet-dominated'
regime and the NSs are, ``if not jet-dominated, close to the
transition to this regime.''  A key uncertainty in this suggestion is
whether the scaling relation that has been established for BHs between
radio luminosity and jet luminosity ($L_{\rm radio} \propto L_{\rm
x}^{0.7}$) is also valid for NSs, as is assumed.
In addition, Occam's razor suggests that the relative faintness of BHs
is unlikely to be attributable to their stronger jets: In \S4.5, we
show that Sagittarius A$^*$, which is radiating in quiescence at the
same level of Eddington-scaled luminosity as A0620-00, does not
possess an energetically-important jet.

{\it Coronal emission from BH secondaries:} The quiescent X-ray
luminosity of the BH systems has been attributed to a
rotationally-enhanced stellar corona in the secondary star (Bildsten
and Rutledge 2000, but see Lasota 2000).  However, as discussed in
detail by Narayan et al. (2002a), the luminosities of three of the
systems plotted in Figures 8 and 9 exceed by a factor of 6--60 the
maximum predicted luminosity of the coronal model; likewise the three
systems with adequate data quality show X-ray spectra that are
harder/hotter than that typically seen in stellar coronae.  Finally,
if stellar coronae do contribute at some level, then the accretion
luminosities of the BHs are even lower than our estimates, which would
further strengthen the evidence for event horizons.

{\it Incandescent neutron stars:} The quiescent luminosity of NS
transients has been attributed to heating of the star's crust during
outburst followed by cooling in quiescence.  This model likewise has
problems.  (For details and references, see Narayan et al. 2002a.)
Briefly, the rapid variability of the prototypical NS transient system
Cen X-4 is not expected in a cooling model and implies that no more
than about a third of the quiescent luminosity is due to crustal
cooling.  Furthermore, power-law tails that carry about half the total
luminosity are observed for many NS transients (e.g., Cen X-4 and Aql
X-1).  These are unlikely to arise from a cooling NS surface, but
could easily be produced by accretion.  Finally, strong evidence for
continued accretion in quiescence comes from optical variability,
which is widely observed for these quiescent systems.

{\it Pulsar wind/shock emission:} In another accretionless model, the
NS transient switches to a radio pulsar-like mode in quiescence
(Campana \& Stella 2000).  The X-ray luminosity is expected to be of
order the ``pulsar shock'' luminosity $\sim 2 \times 10^{32} ~{\rm
erg\,s^{-1}}$, which is close to the observed level.  Both the
observed power-law and thermal components of emission (see the
previous paragraph) are naturally explained by this model: the former
component is produced by the shock and the latter is radiated from the
NS surface.  The lack of any significant periodicity in the quiescent
emission (in any electromagnetic band) could be a problem for this
model.  Millisecond X-ray pulsars such as SAX J1808.4-3658 do show
periodicities in outburst, but that emission is clearly the result of
accretion, not a pulsar wind/shock.

{\it Optical/UV luminosity:} Campana \& Stella (2000) note that the
comparison shown in Figs. 8 and 9 assumes the X-ray luminosity is an
accurate measure of the accretion rate near the BH or NS.  They argue
that the optical and UV luminosity also originates near the central
object and should therefore be included in the comparison.  The
non-stellar optical/UV luminosity is much greater than the X-ray
luminosity.  When it is included, the difference between the BH and NS
systems largely disappears.  However, as detailed in Narayan et
al. (2002a), there are some problems with this argument.  The level of
optical/UV emission generated in the inner region of ADAFs is strongly
suppressed by winds and convection, as evidenced by a comparison of
the luminosities of NSs and white dwarfs (Loeb, Narayan \& Raymond
2001).  It thus appears unlikely that the optical and UV emission is
generated in the hot gas close to the accretor (see Shahbaz et
al. 2005).  In the case of CVs, it has been established that a large
fraction of the optical/UV emission comes from the ``hot spot,'' and
it is quite reasonable to expect that this is true for BH and NS
systems as well (there is some evidence to support this; Narayan et
al. 2002a).
Further observations are needed to determine the origin of optical/UV
emission in these systems.

\subsection{Further Evidence for the Event Horizon in BHBs}

In addition to the arguments discussed above, we briefly summarize
three additional lines of evidence for the existence of event
horizons.  All three are based on comparisons between BH and NS X-ray
binaries.  As in the examples above, the first argument considers
quiescent systems, whereas the latter two consider active states of
accretion.

In quiescence, a soft component of thermal emission is very commonly
observed from the surfaces of accreting NSs, which is widely
attributed to either deep crustal heating (Brown, Bildsten \& Rutledge
1998) or to accretion.  No such component is present in the spectrum
of the quiescent BHB XTE J1118+480 (hereafter J1118), as one would
expect if the compact X-ray source is a bona fide BH that possesses an
event horizon (McClintock et al.\ 2004).  Because of the remarkably
low column density to J1118 ($N_{\rm H} \approx 1.2 \times
10^{20}$~cm$^{-2}$) the limit on a hypothetical thermal source is very
strong ($kT_{\infty}<$ 0.011 keV); it is in fact a factor of $\sim 25$
lower in flux than the emission predicted by the theory of deep
crustal heating, assuming that J1118 has a material surface analogous
to that of NSs.  Likewise, there is no evidence that accretion is
occurring in quiescence onto the surface of J1118, which is the
mechanism often invoked to explain the far greater thermal
luminosities of NSs.  The simplest explanation for the absence of any
thermal emission is that J1118 lacks a material surface and possesses
an event horizon.

Type I X-ray bursts are very common in NS X-ray binaries, but no type
I burst has been seen among the BH systems.  A model developed by
Narayan \& Heyl (2002, 2003), which reproduces the gross observational
trends of bursts in NS systems, shows that, if the dynamical BHs have
surfaces, they should exhibit instabilities similar to those that lead
to type I bursts on NSs.  Remillard et al.\ (2006), following earlier
work by Tournear et al. (2003), measured the rates of type I X-ray
bursts from a sample of 37 nonpulsing X-ray transients observed with
{\it RXTE} during 1996--2004.  Among the NS sources, they found 135
type I bursts in 3.7 Ms of PCA exposures (13 sources) with a burst
rate function consistent with the Narayan \& Heyl model.  However, for
the BH group (18 sources), they found no confirmed type I bursts in
6.5 Ms of exposure.  Their upper limit on the incidence of burst
activity in these sources is inconsistent with the model predictions
at a high statistical significance if the accretors in BHBs have solid
surfaces.  The results provide strong indirect evidence for BH event
horizons, and it would appear that the evidence can be refuted only by
invoking rather exotic physics.

Likewise drawing on the extensive archive of {\it RXTE} data, Done and
Gierlinski (2003) have examined the patterns of X-ray spectral
evolution of active BH and NS sources and identified a distinct type
of soft spectrum that is occasionally observed only in the BH sources.
They attribute this spectrum to thermal emission from the inner
accretion disk (corresponding to the high state discussed earlier).
They then argue that NSs with a similar accretion rate cannot exhibit
such a simple, low-temperature spectrum because they would have a
second component in the emission from the boundary layer where
accreting matter impacts the stellar surface.  They present a
thermal/nonthermal Comptonization model for the boundary layer
emission which has significant uncertainty, given that so many details
of the accretion physics are complex and poorly understood.
Nevertheless, Done \& Gierlinski (2003) do appear to have identified a
systematic difference in the X-ray spectra of accreting BHs and NSs,
and it is surely worth pursuing this signature in the effort to amass
evidence for the reality of event horizons.


\subsection{Event Horizon in Sagittarius A$^*$}

The supermassive BH in Sgr A$^*$ has a mass $M\sim4\times10^6 M_\odot$
(Sch\"odel et al. 2002; Ghez et al. 2005a) but an accretion luminosity
of only $\sim 10^{36} ~{\rm erg\,s^{-1}}$; thus the source is highly
sub-Eddington, $L\ \lta\ 10^{-8}L_{\rm Edd}$.  From our earlier
discussion, the accretion flow must be in the form of an ADAF, i.e.,
the radiating gas must be extremely hot and optically thin.  This
expectation is confirmed by the $10^{10}$ K brightness temperature of
the radio/millimeter emission (Shen et al. 2005) and the fact that
most of the emission is in this band rather than at frequencies $\nu
\sim kT/h$ (X-ray/$\gamma$-ray band).

The ADAF model has been successfully applied to Sgr A$^*$ (Narayan et
al. 1995, 1998a; Manmoto et al. 1997; Mahadevan 1998; Yuan et al. 2003;
to mention a few).  If we assume that the accretion flow is
radiatively very inefficient, then it is easy to make a strong case
for Sgr A$^*$ not having a surface (Narayan et al. 1998a).  However, as
Broderick \& Narayan (2006, 2007) showed, we can argue for the
presence of an event horizon even without assuming an ADAF.

If Sgr A$^*$ does not have an event horizon, but has a surface, then
any emission from the surface will be blackbody-like.  This is because
we expect the surface to be optically thick.  However, the
radio/millimeter emission mentioned above cannot be from this surface
because of its incredibly high $10^{10}$ K brightness temperature.
Thus, the radio/millimeter radiation is from the accretion flow, and
its luminosity gives a strict lower bound on $L_{\rm acc}$.  By
equation (\ref{Lsurfacc}) then, we expect a luminosity of at least
$L_{\rm surf}\sim 10^{36} ~{\rm erg\,s^{-1}}$ from the surface.  The
surface radiation would be thermal and blackbody-like and should come
out in the infrared (as easily shown, given the luminosity and the
likely area of the surface).  However, there is no sign of it!

\begin{figure}
\includegraphics[width=3.3in,clip]{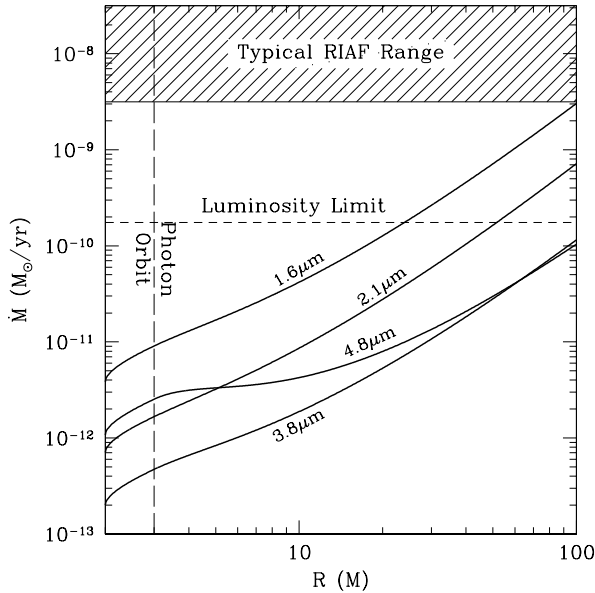}

\caption{The four solid lines show independent upper limits on the
mass accretion rate at the surface of Sgr A$^*$ (assuming the source
has a surface) as a function of the surface radius $R$.  Each limiting
curve is derived from a limit on the quiescent flux of Sgr A$^*$ in an
infrared band.  The hatched area at the top labeled ``Typical RIAF
Range'' corresponds to the mass accretion rate in typical ADAF models
of Sgr A$^*$ (e.g., Yuan et al. 2003).  The horizontal dashed line
represents the minimum accretion rate needed to power the bolometric
luminosity of Sgr A$^*$.  (From Broderick \& Narayan 2006)}
\end{figure}

Figure 10 shows constraints on the mass accretion rate in Sgr A$^*$.
The horizontal dashed line is the minimum accretion rate $\dot{M}_{\rm
acc}$ in Sgr A$^*$; this is the rate needed, even with a radiatively
efficient flow ($\eta\sim0.1$), just to power the observed
radio/millimeter radiation.  The four solid lines show the maximum
mass accretion rate $\dot{M}_{\rm surf}$ allowed if Sgr A$^*$ has a
spherical surface of radius $R$.  Each curve corresponds to a measured
limit on the steady quiescent infrared flux in a particular band
(Stolovy et al. 2003; Clenet et al. 2004; Ghez et al. 2005b) and
provides an independent upper limit on $\dot{M}_{\rm surf}$.  For
radii less than $15GM/c^2$ (the upper limit on the size of Sgr A$^*$;
Shen et al. 2005), we see that all four bands give upper limits on
$\dot{M}_{\rm surf}$ that are far below the minimum $\dot{M}_{\rm
acc}$; in fact, the discrepancy is larger than a factor of 100 in the
$3.8 ~\mu{\rm m}$ band.  Note that the discrepancy would be much
larger if we assumed that the accretion flow is radiatively
inefficient (corresponding to the hatched region in Fig. 10).

The only way to avoid the large discrepancy illustrated in Fig. 10 is
to give up the assumption that Sgr A$^*$ has a surface.  Obviously, if
the source has an event horizon, then we do not expect any surface
radiation, and there is no problem.

What if Sgr A$^*$ ejects all the accreting mass via a jet before the
gas reaches the surface?  Is this a viable explanation for the lack of
surface emission?  We can easily rule out this possibility.  Recall
that the source of energy in an accretion system is gravity.  At a
bare minimum, we know that matter is being accreted at a rate
$\dot{M}_{\rm acc}$ (shown by the horizontal dashed line in Fig. 10)
in order to produce the observed radiation.  All of this mass {\it
has} to fall into the potential well in order to release its energy.
If Sgr A$^*$ has a jet with a kinetic luminosity $L_{\rm jet}$, the
energy for this must also come from accretion.  We will then require
an even larger $\dot{M}_{\rm acc}$, and hence a larger $\dot{M}_{\rm
surf}$, since the accreting gas now has to power both the radiation
and the jet.  The discrepancy with the observed limits on the
quiescent infrared flux would then be even larger.

\subsection{Effects of Very Strong Gravity}

The arguments for the event horizon presented in \S\S4.2, 4.5 are
based on Newtonian ideas.  Some authors (e.g., Abramowicz, Kluzniak \&
Lasota 2002) have questioned whether the arguments might be
substantially modified by strong gravity in the vicinity of the
compact accretor.

Buchdahl (1959) showed, for a wide class of reasonable equations of
state, that the smallest radius allowed for a compact non-rotating
object is $R_{\rm min}= (9/8)R_S$.  An object with this limiting
radius has a gravitational redshift from its surface $z_{\rm gr}=2$.
At such modest redshifts, we do not expect the effects of strong
gravity to be particularly large, and so the arguments presented in
\S\S4.2, 4.5 will continue to hold.

Recently, however, a new class of solutions has been discussed, which
goes variously under the name of ``gravastar'' and ``dark energy
star'' (Mazur \& Mottola 2001; Chapline et al. 2003; Visser \&
Wiltshire 2004; Carter 2005; Lobo 2006).  In this model, the radius
$R$ of an object of mass $M$ is allowed to be arbitrarily close to
$R_S$: $\Delta R \equiv R-R_S \ll R_S$.  The surface redshift can
then be arbitrarily large:
\begin{equation}
1+z_{\rm gr}\approx (R_S/\Delta R)^{1/2} \gg 1.
\end{equation}
Although the gravastar model is very artificial\footnote{Broderick \&
Narayan first submitted their 2007 paper to Phys. Rev. Lett., but the
paper was not even sent out for review because, in the words of the
editor, ``the gravastar model of Mazur and Mottola is not considered
by the community to be a viable alternative.''  B\&N then submitted
the paper to Class. Quantum Grav., where most papers on gravastars and
dark energy stars are published.}, it is nevertheless interesting to
ask whether such a model, which has no event horizon, can explain the
observations described in \S\S4.2, 4.5 (Abramowicz et al. 2002).
Broderick \& Narayan (2006, 2007) show that it cannot.  We briefly
summarize the arguments here.

A large value of $z_{\rm gr}$ means that any radiation emitted from
the surface is redshifted greatly before it reaches the observer at
infinity.  This looks like an easy way of hiding the surface
luminosity.  However, a simple energy conservation argument shows
otherwise.  Assuming steady state, in the limit as $\Delta R/R_S\to
0$, the total luminosity observed at infinity, $L_{\rm acc} +L_{\rm
surf}$, {\it must} be equal to $\dot{M}c^2$ (the rate of accretion of
rest mass energy).  Moreover, the effective radius of the source as
viewed by a distant observer is $(3\sqrt{3}/2)R_S=2.6R_S$, and so the
surface radiation will be in the X-ray band for accreting BHBs
(\S4.2) and in the infrared for Sgr A$^*$ (\S4.5), exactly as in the
Newtonian analysis.

Could the large gravitational redshift cause a large delay in the
signals from the surface of the star, and could this be why we do not
see the radiation?  The extra delay due to relativity is easily
estimated by considering null geodesics in the Schwarzschild metric:
\begin{equation}
\Delta t = (R_S/c)\ln(R_S/\Delta R).
\end{equation}
Since the dependence is only logarithmic, the extra delay is not
significant.  For instance, even if we take $\Delta R$ to be
comparable to the Planck length $\sim 10^{-33}$ cm (the smallest
length we can legitimately consider), the delay is only $\sim 10$ ms
for a BHB and $\sim 1000$ s for Sgr A$^*$.

We assumed that the radiation from the surface would have a
blackbody-like spectrum.  Is this likely?  Actually, it is virtually
guaranteed as $\Delta R/R_S\to0$.  When the radius of an object is
very close to the Schwarzschild radius, most rays emitted from the
surface are bent back on to the surface, and only a tiny fraction of
rays escapes to infinity.  The solid angle corresponding to escape is
$\Omega_{\rm esc} \sim \Delta R/R_S \ll 1$.  In this limit, the
surface behaves just like a furnace with a pinhole (the textbook
example of a blackbody!), and so the escaping radiation is certain to
have a nearly perfect blackbody spectrum.

Could the surface luminosity escape in the form of particles rather
than radiation?  By the blackbody argument given above, whatever
escapes must be in thermodynamic equilibrium at the temperature
$T_\infty$ observed at infinity.  For the sources of interest to us,
$kT_\infty \sim$ eV (Sgr A$^*$) and keV (BHBs).  In thermodynamic
equilibrium, the only particles with any significant number density at
these temperatures are neutrinos.  Even if we include all three
species of neutrinos, the radiation flux is reduced by a factor of
only 8/29 (Broderick \& Narayan 2007).  This is a small correction
compared to the large discrepancies $\gta\ 100$ that we described in
\S\S4.2, 4.5.

Finally, we note that our arguments for the event horizon are based on
a steady state assumption.  Specifically, we assume that the surface
luminosity is proportional to the average mass accretion rate on the
surface.  In the gravastar model, in particular, one can imagine
scenarios in which steady state is not reached.  However, Broderick \&
Narayan (2007) show that this explanation can be ruled out, at least
with current gravastar models.

We thus conclude that strong gravity effects are unable to weaken our
arguments for the presence of an event horizon in quiescent BHBs and
in Sgr A$^*$.  We must look to more conventional astrophysical
explanations.  
The only idea that we find somewhat plausible is that the radiation we
observe from quiescent X-ray binaries and Sgr A$^*$ is not produced by
accretion at all, but by some other process.  This would undercut all
our arguments since we assume, as an article of faith, that the
radiation we observe is powered by gravity through accretion.  Given
the success of the accretion paradigm in explaining a vast body of
observations on a variety of BH systems in many different spectral
states, it seems rather extreme to abandon the idea of accretion in
just those particular sources where we find evidence for the presence
of an event horizon.


\section{Conclusion}

The aim of this article is two-fold: To give an updated and current
account of the ADAF model, with an emphasis on applications to
accreting BHs (\S3), and to review the considerable body of evidence
presently available for the presence of event horizons in
astrophysical BHs (\S4).

Observations of BH binaries (BHBs) and active galactic nuclei (AGN)
indicate that, at luminosities of a few percent or less of Eddington,
accretion occurs via a very different mode than the standard thin
accretion disk.  At these luminosities, sources have hard X-ray
spectra, quite unlike the soft blackbody-like spectra seen in more
luminous sources.  Prior to the establishment of the ADAF model,
observations in the hard state were modeled in an ad hoc way using
empirical thermal Comptonization models.  In the mid-1990s, the ADAF
solution was shown to have precisely the densities, temperatures,
radiative (in)efficiencies -- and stability -- required to provide a
physical description of the observations.  Moreover, since the ADAF
model is essentially mass-independent, observations of BHBs and AGN
are both explained with more-or-less the same ADAF model.  The model
gives satisfactory results for a wide range of luminosity, from about
$10^{-1}$ down to $10^{-8}$ of Eddington (below which there are no
observations).

The defining characteristic of an ADAF is that the gas is not
radiatively efficient.  A significant fraction of the energy released
by viscous dissipation is retained in the accreting gas and advected
with the flow.  The trapped thermal energy causes the accreting gas to
be weakly bound to the BH.  Based on this property, it was predicted
already in the earliest ADAF papers that sources in the ADAF state
would have strong winds and jets.  Nonthermal radio emission has been
detected in recent years from BHBs in the hard state and quiescent
state, as well as from all AGN with luminosities below $\sim1\%$ of
Eddington.  These sources have spectra consistent with the ADAF model,
thus confirming a strong connection between ADAFs and jets/outflows.
A currently active topic of research is the role of AGN feedback on
galaxy formation.  Research in this area can now draw on the extensive
literature on ADAFs.

Of special interest in this work are those accreting objects for which
there exist estimates or constraints on the two system specific
parameters: BH mass $M$ and mass accretion rate $\dot M$.
Specifically, we are referring here to Saggitarius A$^*$ and a
selected sample of quiescent BH and NS X-ray binaries.  We have
featured these two examples because they provide strong observational
evidence for the existence of the event horizon.

In the case of the binaries, one can make a plausible argument that,
for comparable orbital periods, the mass accretion rates and
luminosities of both types of systems should be comparable.
Remarkably, however, the BH systems are observed to be dimmer by a
factor of $\sim100-1000$.  We review a wide range of unsatisfactory
attempts to explain this large luminosity difference.  In contrast,
the ADAF model provides an entirely straightforward explanation for
the faintness of quiescent BHs: A NS must radiate the trapped thermal
radiation that is advected with the accretion flow and rains down on
its surface, whereas the BH hides the energy behind its event horizon.
Additional evidence for the event horizon is provided in further
comparisons of NSs and BHs in outburst: The BHs lack type I X-ray
bursts, and they lack a distinctive boundary-layer component of
emission.  Both of these properties are expected if the BHs possess
event horizons, but hard to explain otherwise.

The supermassive BH in Sgr A$^*$ is extraordinarily quiescent.  The
observations strongly support the existence of an ADAF, and it is
consequently easy to provide a compelling argument for an event
horizon.  However, for Sgr A$^*$, one can make an even stronger
argument for the lack of a hypothetical material surface without at
all invoking an ADAF.  The large radio/millimeter accretion luminosity
of Sgr A$^*$, which has a brightness temperature $>10^{10}$ K,
obviously cannot be emitted by an optically-thick surface, and so it
must be radiated from the accretion flow.  This establishes a hard
lower limit on the accretion luminosity and $\dot{M}$.  Meanwhile,
high angular resolution radio observations constrain the radius of the
surface to be $< 15R_{\rm S}$.  This constraint and the lower limit on
$\dot M$ predict a near-IR flux of thermal, blackbody-like surface
emission that is far above the observed limits.  The obvious
explanation is that there is no material surface, only an event
horizon.  As an added bonus, it is energetically impossible to explain
away the lack of surface emission by appealing to mass loss in a jet.

The case for event horizons in both Sgr A$^*$ and in the stellar-mass
BHs is robust against appeals to strong gravity or the leading models
of exotic stars.  First, GR effects will be mild for nearly all
conventional models of degenerate stars, whose radii are restricted to
be $\ge 9/8R_{\rm S}$ and whose surface redshifts are therefore $\le
2$.  Secondly, even for an exotic star (a gravastar) with an
extraordinary surface redshift of a million or more, the full
accretion luminosity from its surface will be delivered as X-rays (in
binaries) or infrared (in Sgr A$^*$) to a distant observer.  Thus,
extreme redshifts have practically no effect on the argument for the
event horizon.

To play on Carl Sagan's famous comment, there will always be an
absence of {\it direct} evidence for the event horizon, but this
surely cannot be taken as evidence of its absence in nature.  On the
contrary, and with a pun in mind, many indicators show that the event
horizon is an inescapable reality.

\medskip\noindent The authors thank Bozena Czerny, Jean-Pierre Lasota
and Feng Yuan for helpful comments.





\begin{thebibliography}{}


\bibitem[Abramowicz et al.(1996)]{abr96}
Abramowicz, M. A., Chen, X., Granath, M., \& Lasota, J. P. 1996,
ApJ, 471, 762

\bibitem[Abramowicz et al.(1995)]{abr95}
Abramowicz, M. A., Chen, X., Kato, S., Lasota, J. P., \& Regev, O.
1995, ApJ, 438, L37

\bibitem[Abramowicz et al.(1988)]{abr88} 
Abramowicz, M. A., Czerny, B., Lasota, J. P., \& Szuszkiewicz,
E. 1988, ApJ, 332, 646

\bibitem[Abramowicz et al.(2002)]{abr02}
Abramowicz, M. A., Kluzniak, W., \& Lasota, J. P. 2002, A\&A, 396, L31

\bibitem[Abramowicz et al.(2000)]{ali00}
Abramowicz, M. A., Lasota, J. P., \& Igumenshchev, I. V.
2000, MNRAS, 314, 775

\bibitem[Arnaud(1996)]{arn96}
Arnaud, K. A. 1996, in ASP Conf. Ser. 191, Astronomical Data Analysis and
Systems V, ed. G. H. Jacoby \& J. Barnes (San Francisco: ASP) p17

\bibitem[Baganoff(2001)]{bag01}
Baganoff, F. K., et al. 2001, Nature, 413, 45

\bibitem[Baganoff(2003)]{bag03}
Baganoff, F. K., et al. 2003, ApJ, 591, 891

\bibitem[Balbus \& Hawley(2002)]{bal02}
Balbus, S. A., \& Hawley, J. F. 2002, ApJ, 573, 749

\bibitem[Balucinska-Church et al.(1995)]{bal95}
Balucinska-Church, M., Belloni, T., Church, M. J., \&
Hasinger, G. 1995, A\&A, 302, L5

\bibitem[Baum et al.(1995)]{bau95}
Baum, S. A., Zirbel, E. L., \& O'Dea, C. P. 1995, ApJ, 451, 88

\bibitem[Begelman(1979)]{beg79}
Begelman, M. C. 1979, MNRAS, 187, 237

\bibitem[Begelman \& Celotti(2004)]{beg04}
Begelman, M. C., \& Celotti, A. 2004, MNRAS, 352, L45

\bibitem[Begelman \& Chiueh(1988)]{bc88}
Begelman, M. C., \& Chiueh, T. 1988, ApJ, 332, 872

\bibitem[Begelman \& Meier(1982)]{beg82}
Begelman, M. C., \& Meier, D. L. 1982, ApJ, 253, 873

\bibitem[Bildsten \& Rutledge(2000)]{bil00}
Bildsten, L., \& Rutledge, R. E. 2000, ApJ, 541, 908

\bibitem[Bisnovatyi-Kogan \& Lovelace(1997)]{bl97}
Bisnovatyi-Kogan, G., \& Lovelace, R. V. E. 1997, ApJ, 486, L43

\bibitem[Blackman(1999)]{bla99}
Blackman, E. G. 1999, MNRAS, 302, 723

\bibitem[Blandford \& Begelman(1999)]{bb99}
Blandford, R. D., \& Begelman, M. C. 1999, MNRAS, 303, L1

\bibitem[Broderick \& Narayan(2006)]{bn06}
Broderick, A. E., \& Narayan, R. 2006, ApJ, 638, L21

\bibitem[Broderick \& Narayan(2007)]{bn07}
Broderick, A. E., \& Narayan, R. 2007, Class. Quantum Grav., 24, 659

\bibitem[Brown et al.(1998)]{bbr98}
Brown, E. F., Bildsten, L., \& Rutledge, R. E. 1998, ApJ, 504, L95

\bibitem[Buchdahl(1959)]{buc59}
Buchdahl, H. A. 1959, Phys. Rev., 116, 1027

\bibitem[Campana \& Stella(2000)]{cam00}
Campana, S., \& Stella, L. 2000, ApJ, 541, 849

\bibitem[Carter(2005)]{car05}
Carter, B. M. N. 2005, Class. Quantum Grav., 22, 4551

\bibitem[Chapline et al.(2003)]{cha03}
Chapline, G., et al. 2003, Int. J. Mod. Phys. A, 18, 3587

\bibitem[Chen et al.(1997)]{che97}
Chen, X., Abramowicz, M. A., \& Lasota, J. P. 1997, ApJ, 476, 61

\bibitem[Chen et al.(1995)]{che95}
Chen, X., Abramowicz, M. A., Lasota, J. P., Narayan, R., \& Yi, I. 1995,
ApJ, 443, L61

\bibitem[Chiang \& Blaes(2003)]{chi03}
Chiang, J., \& Blaes, O. 2003, ApJ, 586, 97

\bibitem[Clenet et al.(2004)]{cle04}
Clenet, Y., et al. 2004, A\&A, 424, L21

\bibitem[Corbel et al.(2000)]{cor00}
Corbel, S., et al. 2000, A\&A, 359, 251

\bibitem[Corbel et al.(2003)]{cor03}
Corbel, S., Nowak, M. A., Fender, R. P., Tzioumis, A. K., \& Markoff, S.
2003, A\&A, 400, 1007

\bibitem[Corbel et al.(2006)]{cor06}
Corbel, S., Tomsick, J. A., \& Kaaret, P. 2006, ApJ, 636, 971

\bibitem[Croton et al.(2006)]{cro06}
Croton, D. J., et al. 2006, MNRAS, 365, 11

\bibitem[Cui et al.(1999)]{cui99}
Cui, W., Zhang, S. N., Chen, W., \& Morgan, E. H. 1999, ApJ, 512, L43

\bibitem[D'Angelo et al.(2008)]{dan08}
D'Angelo, C., Giannios, D., Dullemond, C., \& Spruit, H. 2008, 
A\&A, submitted

\bibitem[Davis et al.(2006)]{ddb06}
Davis, S. W., Done, C., \& Blaes, O. M. 2006, ApJ, 647, 525

\bibitem[Davis \& Hubeny(2006)]{dav06}
Davis, S. W., \& Hubeny, I. 2006, ApJS, 164, 530

\bibitem[Deufel et al.(2002)]{deu02}
Deufel, B., Dullemond, C. P., \& Spruit, H. C. 2002, A\&A, 387, 907


\bibitem[Di Matteo et al.(2003)]{dim03}
Di Matteo, T., Allen, S. W., Fabian, A. C., Wilson, A. S., \& Young, A. J.
2003, ApJ, 582, 133

\bibitem[Di Matteo et al.(2000)]{dim00}
Di Matteo, T., Quataert, E., Allen, S. W., Fabian, A. C., \& Narayan, R.
2000, MNRAS, 311, 507

\bibitem[Di Matteo et al.(2005)]{dim05}
Di Matteo, T., Springel, V., \& Hernquist, L. 2005, Nature, 433, 604

\bibitem[Di Salvo et al.(2001)]{dis01}
Di Salvo, T., Done, C., Zycki, P. T., Burderi, L., \& Robba, N. R. 2001,
ApJ, 547, 1024

\bibitem[Done \& Gierlinski(2003)]{dg03}
Done, C., \& Gierlinski, M. 2003, MNRAS, 342, 1041

\bibitem[Done et al.(2007)]{don07}
Done, C., Gierlinski, M., \& Kubota, A. 2007, A\&AR, 15, 1

\bibitem[Dubus et al.(2001)]{dub01}
Dubus, G., Hameury, J. M., \& Lasota, J. P. 2001, A\&A, 373, 251

\bibitem[Dullemond \& Spruit(2005)]{dul05}
Dullemond, C. P., \& Spruit, H. C. 2005, A\&A, 434, 415

\bibitem[Esin et al.(1997)]{esi97}
Esin, A. A., McClintock, J. E., \& Narayan, R. 1997, ApJ, 489, 865

\bibitem[Esin et al.(1998)]{esi98}
Esin, A. A., Narayan, R., Cui, W., Grove, J. E., \& Zhang, S. N. 1998, 
ApJ, 505, 854

\bibitem[Esin et al.(2001)]{esi01}
Esin, A. A., et al. 2001, ApJ, 555, 483

\bibitem[Fabian \& Rees(1995)]{fab95}
Fabian, A. C., \& Rees, M. J. 1995, MNRAS, 277, L5

\bibitem[Falcke et al.(1998)]{fal98}
Falcke, H., Goss, W. M., Matsuo, H., Teuben, P., Zhao, J. H., \& Zylka,
R. 1998, ApJ, 499, 731

\bibitem[Falcke et al.(2000)]{fal00}
Falcke, H., Nagar, N. M., Wilson, A. S., \& Ulvestad, J. S. 2000,
ApJ, 542, 197

\bibitem[Fender(2001)]{fen01}
Fender, R. P. 2001, MNRAS, 322, 31

\bibitem[Fender \& Belloni(2004)]{fb04}
Fender, R. P., \& Belloni, T. 2004, ARAA, 42, 317

\bibitem[Fender et al.(2004)]{fbg04}
Fender, R. P., Belloni, T. M., \& Gallo, E. 2004, MNRAS, 355, 1105

\bibitem[Fender et al.(2003)]{fen03}
Fender, R. P., Gallo, E., \& Jonker, P. G. 2003, MNRAS, 343, L99

\bibitem[Frank et al.(2002)]{fkr02}
Frank, J., King, A., \& Raine, D. J. 2002, Accretion Power in Astrophysics
(Cambridge Univ. Press)

\bibitem[Gallo et al.(2005)]{gal05}
Gallo, E., Fender, R. P., \& Hynes, R. I. 2005, MNRAS, 356, 1017

\bibitem[Gallo et al.(2003)]{gal03}
Gallo, E., Fender, R. P., \& Pooley, G. G. 2003, MNRAS, 344, 60

\bibitem[Gallo et al.(2006)]{gal06}
Gallo, E., et al. 2006, MNRAS, 370, 1351

\bibitem[Gammie et al.(1999)]{gam99}
Gammie, C. F., Narayan, R., \& Blandford, R. 1999, ApJ, 516, 177

\bibitem[Garcia et al.(1998)]{gar98} 
Garcia, M. R., McClintock, J. E.,
\& Narayan, R. 1998, ASP Conf. Series, Vol. 137, ed. S. Howell,
E. Kuulkers \& C. Woodward, p.506

\bibitem[Garcia et al.(2001)]{gar01}
Garcia, M. R., McClintock, J. E., Narayan, R., Callanan, P.,
Barret, D., \& Murray, S. S. 2001, ApJ, 553, L47

\bibitem[Ghez et al.(2005a)]{ghe05a}
Ghez, A. M., et al. 2005a, ApJ, 620, 744

\bibitem[Ghez et al.(2005b)]{ghe05b}
Ghez, A. M., et al. 2005b, ApJ, 635, 1087

\bibitem[Gierlinski et al.(1997)]{gie97}
Gierlinski, M., et al. 1997, MNRAS, 288, 958

\bibitem[Gilfanov et al.(1999)]{gil99}
Gilfanov, M., Churazov, E., \& Revnivtsev, M. 1999, A\&A, 352, 182

\bibitem[Goldston et al.(2005)]{gol05}
Goldston, J. E., Quataert, E., \& Igumenshchev, I. V. 2005, ApJ, 621, 785

\bibitem[Hameury et al.(1999)]{ham99}
Hameury, J. M., Lasota, J. P., \& Dubus, G. 1999, MNRAS, 303, 39

\bibitem[Hameury et al.(1997)]{ham97}
Hameury, J. M., Lasota, J. P., McClintock, J. E., \& Narayan, R.
1997, ApJ, 489, 234

\bibitem[Hameury et al.(2003)]{ham03}
Hameury, J. M., et al. 2003, A\&A, 399, 631

\bibitem[Hameury et al.(2007)]{ham07}
Hameury, J. M., Lasota, J. P., \& Viallet, M. 2007, in Black Holes from
Stars to Galaxies, ed. V. Karas \& G. Matt, Cambridge Univ. Press, p297

\bibitem[Hawley \& Balbus(2002)]{haw02}
Hawley, J. F., \& Balbus, S. A. 2002, ApJ, 573, 738

\bibitem[Hawley et al.(1996)]{haw96}
Hawley, J. F., Gammie, C. F., \& Balbus, S. A. 1996, ApJ, 464, 690

\bibitem[Heinz(2004)]{hei04}
Heinz, S. 2004, 355, 835

\bibitem[Heinz et al.(2005)]{hei05}
Heinz, S., Merloni, A., Di Matteo, T., \& Sunyaev, R. 2005, Astrophys.
Sp. Sci., 300, 15

\bibitem[Heinz \& Sunyaev(2003)]{hei03}
Heins, S., \& Sunyaev, R. 2003, MNRAS, 343, L59

\bibitem[Hjellming et al.(2000)]{hje00}
Hjellming, R. M., Rupen, M. P., Mioduszewski, A. J., \& Narayan, R.
2000, ATel, \#54

\bibitem[Ho(1999)]{ho99}
Ho, L. C. 1999, ApJ, 516, 672

\bibitem[Ho(2002)]{ho02}
Ho, L. C. 2002, ApJ, 564, 120

\bibitem[Honma(1996)]{hon96}
Honma, F. 1996, PASJ, 48, 77

\bibitem[Hopkins et al.(2006a)]{hop06a}
Hopkins, P. F., et al. 2006a, ApJS, 163, 1

\bibitem[Hopkins et al.(2006b)]{hop06b}
Hopkins, P. F., Narayan, R., \& Hernquist, L. 2006b, ApJ, 643, 641

\bibitem[Hornstein et al.(2002)]{hor02}
Hornstein, S. D., Ghez, A. M., Tanner, A., Morris, M., Becklin, E. E.,
\& Wizinowich, P. 2002, ApJ, 577, 738

\bibitem[Hynes et al.(2003)]{hyn03}
Hynes, R. I., Charles, P. A., Casares, J., Haswell, C. A., Zurita, C.,
\& Shahbaz, T. 2003, MNRAS, 340, 447

\bibitem[Ichimaru(1977)]{ich77}
Ichimaru, S. 1977, ApJ, 214, 840

\bibitem[Igumenshchev(2004)]{igu04}
Igumenshchev, I. V. 2004, Prog. Theor. Phys. Suppl., 155, 87

\bibitem[Igumenshchev(2006)]{igu06}
Igumenshchev, I. V. 2006, ApJ, 649, 361

\bibitem[Igumenshchev \& Abramowicz(2000)]{ia00}
Igumenshchev, I. V., \& Abramowicz, M. A. 2000, ApJS, 130, 463

\bibitem[Igumenshchev et al.(2000)]{igu00}
Igumenshchev, I. V., Abramowicz, M. A., \& Narayan, R. 2000,
ApJ, 537, L27

\bibitem[Igumenshchev \& Narayan(2002)]{igu02}
Igumenshchev, I. V., \& Narayan, R. 2002, ApJ, 566, 137

\bibitem[Igumenshchev et al.(2003)]{igu03}
Igumenshchev, I. V., Narayan, R., \& Abramowicz, M. A. 2003,
ApJ, 592, 2042

\bibitem[Jonker et al.(2007)]{jon07}
Jonker, P. G., Steeghs, D., Chakrabarty, D., \& Juett, A. M. 2007,
ApJ, 665, L147

\bibitem[Kato et al.(1998)]{kfm98}
Kato, S., Fukue, J., \& Mineshige, S. 1998, Black Hole Accretion Disks
(Kyoto Univ. Press)

\bibitem[Kato et al.(1997)]{kat97}
Kato, S., Yamasaki, T., Abramowicz, M. A., \& Chen, X. 1997, PASJ, 49, 221

\bibitem[Lasota(1996)]{las96}
Lasota, J. P. 1996, in Physics of Accretion Disks, ed. S. Kato, 
J. Fukue \& S. Mineshige (Gordon \& Breach) p85

\bibitem[Lasota(1999a)]{las99a}
Lasota, J. P. 1999a, Phys. Rep., 311, 247

\bibitem[Lasota(1999b)]{las99b}
Lasota, J. P. 1999b, Sci. Am., 280, 30

\bibitem[Lasota(2000)]{las00}
Lasota, J. P. 2000, A\&A, 360, 575

\bibitem[Lasota(2001)]{las01}
Lasota, J. P. 2001, in Black Hole Binaries and Galactic Nuclei,
ed. L. Kaper, E. P. J. van den Heuvel \& P. A. Woudt, Springer-Verlag, p149

\bibitem[Lasota(2007)]{las07}
Lasota, J. P. 2007, Comptes Rendus Physique, 8, 45

\bibitem[Lasota(2008)]{las08}
Lasota, J. P. 2008, New Astron. Rev., ed. M. A. Abramowicz 
(astro-ph/0801.0490)

\bibitem[Lasota et al.(1996a)]{las96a}
Lasota, J. P., Abramowicz, M. A., Chen, X., Krolik, J., Narayan, R., \&
Yi, I. 1996a, ApJ, 462, 142

\bibitem[Lasota \& Hameury(1998)]{las98}
Lasota, J. P., \& Hameury, J. P. 1998, AIP conf. Series, 431, 351

\bibitem[Lasota et al.(1996b)]{las96b}
Lasota, J. P., Narayan, R., \& Yi, I. 1996b, A\&A, 314, 813

\bibitem[Li et al.(2005)]{li05}
Li, L. X., Zimmerman, E. R., Narayan, R., \& McClintock, J. E.
2005, ApJS, 157, 335

\bibitem[Liu et al.(2007)]{liu07}
Liu, B. F., Taam, R. F., Meyer-Hofmeister, E., \& Meyer, F. 2007,
ApJ, 671, 695

\bibitem[Lobo(2006)]{lob06}
Lobo, F. S. N. 2006, Class. Quantum Grav., 23, 1525

\bibitem[Loeb et al.(2001)]{lnr01}
Loeb, A., Narayan, R., \& Raymond, J. C. 2001, ApJ, 547, L151

\bibitem[Loewenstein et al.(2001)]{loe01}
Loewenstein, M., Mushotzky, R. F., Angelini, L., Arnaud, K., \&
Quataert, E. 2001, ApJ, 555, L21

\bibitem[Maccarone \& Coppi(2003)]{mac03}
Maccarone, T. C., \& Coppi, P. S. 2003, MNRAS, 338, 189

\bibitem[Machida et al.(2001)]{mch01}
Machida, M., Matsumoto, R., \& Mineshige, S. 2001, PASJ, 53, L1

\bibitem[Machida et al.(2004)]{mch04}
Machida, M., Nakamura, K., \& Matsumoto, R. 2004, PASJ, 56, 671

\bibitem[Machida et al.(2006)]{mch06}
Machida, M., Nakamura, K., \& Matsumoto, R. 2006, PASJ, 58, 193

\bibitem[Mahadevan(1997)]{mah97}
Mahadevan, R. 1997, ApJ, 477, 585

\bibitem[Mahadevan(1998)]{mah98}
Mahadevan, R. 1998, Nature, 394, 651

\bibitem[Mahadevan \& Quataert(1997)]{mq97}
Mahadevan, R., \& Quataert, E. 1997, ApJ, 490, 605

\bibitem[Malzac et al.(20040]{mal04}
Malzac, J., Merloni, A., \& Fabian, A. C. 2004, MNRAS, 351, 253

\bibitem[Manmoto(2000)]{man00}
Manmoto, T. 2000, ApJ, 534, 734

\bibitem[Manmoto et al.(1997)]{man97}
Manmoto, T., Mineshige, S., \& Kusunose, M. 1997, ApJ, 489, 791

\bibitem[Maraschi \& Tavecchio(2003)]{mar03}
Maraschi, L., \& Tavecchio, F. 2003, ApJ, 593, 667

\bibitem[Markoff et al.(2001)]{mar01}
Markoff, S., Falcke, H., \& Fender, R. P. 2001, A\&A, 372, L25

\bibitem[Markoff et al.(2005)]{mar05}
Markoff, S., Nowak, M. A., \& Wilms, J. 2005, ApJ, 635, 1203

\bibitem[Mayer \& Pringle(2007)]{may07}
Mayer, M., \& Pringle, J. E. 2007, MNRAS, 376, 435

\bibitem[Mazur \& Mottola(2001)]{maz01}
Mazur, P. O., \& Mottola, E. 2001, gr-qc/0109035

\bibitem[McClintock, J. E. et al.(2004)]{mcc04}
McClintock, J. E., Narayan, R., \& Rybicki, G. B. 2004,
ApJ, 615, 402

\bibitem[McClintock \& Remillard(2007)]{mr07}
McClintock, J. E., \& Remillard, R. A. 2006, in Black Hole Binaries,
ed. W. Lewin \& M. van der Klis (Cambridge Univ. press) p157

\bibitem[McClintock et al.(2006)]{mcc06}
McClintock, J. E., Shafee, R., Narayan, R., Remillard, R. A., Davis,
S. W., \& Li, L. X. 2006, ApJ, 652, 518

\bibitem[McKinney \& Gammie(2004)]{mck04}
McKinney, J. C., \& Gammie, C. F. 2004, ApJ, 611, 977

\bibitem[McKinney(2005)]{mck05}
McKinney, J. C. 2005, ApJ, 630, L5

\bibitem[McKinney(2006)]{mck06}
McKinney, J. C. 2006, MNRAS, 368, 1561

\bibitem[Medvedev(2000)]{med00}
Medvedev, M. V. 2000, ApJ, 541, 811

\bibitem[Meier(2001)]{mei01}
Meier, D. L. 2001, ApJ, 548, L9

\bibitem[Menou et al.(1999a)]{men99a}
Menou, K., Esin, A. A., Narayan, R.,  Garcia, M. R., Lasota, J. P.,
\& McClintock, J. E. 1999a, ApJ, 520, 276

\bibitem[Menou et al.(1999b)]{men99b}
Menou, K., Narayan, R., \& Lasota, J. P. 1999b, ApJ, 513, 811

\bibitem[Menou \& Quataert(2001)]{men01}
Menou, K., \& Quataert, E. 2001, ApJ, 552, 204

\bibitem[Merlone et al.(2003)]{mer03}
Merloni, A., Heinz, S., \& Di Matteo, T. 2003, MNRAS, 345, 1057

\bibitem[Meyer et al.(2000)]{MLM00} 
Meyer, F., Liu, B. F., \& Meyer-Hofmeister, E. 2000, A\&A, 361, 175

\bibitem[Meyer \& Meyer-Hofmeister(1994)]{mm94} 
Meyer, F., \& Meyer-Hofmeister, E. 1994, A\&A, 361, 175

\bibitem[Meyer-Hofmeister et al.(2005)]{MLM05} 
Meyer-Hofmeister, E., Liu, B. F., \& Meyer, F. 2005, A\&A, 432, 181

\bibitem[Miller et al.(2006a)]{mil06a}
Miller, J. M., Homan, J., \& Miniutti, G. 2006a, ApJ, 652, L113

\bibitem[Miller et al.(2006b)]{mil06b}
Miller, J. M., et al. 2006b, ApJ, 653, 525

\bibitem[Miller et al.(2006c)]{mil06c}
Miller, J. M., et al. 2006c, Nature, 441, 953

\bibitem[Mitsuda et al.(1984)]{mit84}
Mitsuda, K., et al. 1984, PASJ, 36, 741

\bibitem[Miyamoto et al.(1995)]{miy95}
Miyamoto, S., et al. 1995, ApJ, 442, L13

\bibitem[Miyoshi et al.(1995)]{mi95}
Miyoshi, M., et al. 1995, Nature, 373, 127

\bibitem[Nagar et al.(2000)]{nag00}
Nagar, N. M., Falcke, H., Wilson, A. S., \& Ho, L. C. 2000,
ApJ, 542, 186

\bibitem[Narayan(1996)]{nar96}
Narayan, R. 1996, ApJ, 461, 136

\bibitem[Narayan(2002)]{nar02}
Narayan, R. 2002, in Lighthouses of the Universe, ed. M. Gilfanov,
R. Sunyaev (Springer) p405

\bibitem[Narayan(2005)]{nar05}
Narayan, R. 2005, Astrophys. Sp. Sci., 300, 177

\bibitem[Narayan et al.(1997a)]{nbm97a}
Narayan, R., Barret, D., \& McClintock, J. E. 1997a, ApJ, 482, 448

\bibitem[Narayan et al.(1997b)]{nar97b}
Narayan, R., Garcia, M. R., \& McClintock, J. E. 1997b, ApJ, 478, L79

\bibitem[Narayan et al.(2002a)]{ngm02a} Narayan, R., Garcia, M. R.,
\& McClintock, J. E. 2002a, in The Ninth Marcel Grossmann Meeting,
ed. V. G. Gurzadyan, R. T. Jantzen \& R. Ruffini, World Scientific, p405

\bibitem[Narayan \& Heyl(2002)]{nh02}
Narayan, R., \& Heyl, J. S. 2002, ApJ, 574, L139

\bibitem[Narayan \& Heyl(2003)]{nh03}
Narayan, R., \& Heyl, J. S. 2003, ApJ, 599, 419

\bibitem[Narayan \& Igumenshchev(2000)]{ni00}
Narayan, R., \& Igumenshchev, I. V. 2000, ApJ, 539, 798

\bibitem[Narayan, Kato \& Honma(1997c)]{nkh97c}
Narayan, R., Kato, S., \& Honma, F. 1997c, ApJ, 476, 49

\bibitem[Narayan et al.(1998a)]{nar98a}
Narayan, R., Mahadevan, R., Grindlay, J. E., Popham, R. G., \&
Gammie, C. 1998a, ApJ, 492, 554

\bibitem[Narayan et al.(1998b)]{nar98b} Narayan, R., Mahadevan, R., \&
Quataert, E. 1998b, in Theory of Black Hole Accretion Disks,
ed. M. A. Abramowicz, G. Bjornsson, \& J. E. Pringle (Cambridge
Univ. Press) p148

\bibitem[Narayan \& McClintock(2005)]{nmc05}
Narayan, R., \& McClintock, J. E. 2005, ApJ, 623, 1017

\bibitem[Narayan et al.(1996)]{nmy96}
Narayan, R., McClintock, J. E., \& Yi, I. 1996, ApJ, 457, 821

\bibitem[Narayan et al.(2002b)]{nqia02b}
Narayan, R., Quataert, E., Igumenshchev, I. V., \& Abramowicz, M. A.
2002b, ApJ, 577, 295

\bibitem[Narayan \& Yi(1994)]{NY94}
Narayan, R., \& Yi, I. 1994, ApJ, 428, L13 (NY94)

\bibitem[Narayan \& Yi(1995a)]{NY95a}
Narayan, R., \& Yi, I. 1995a, ApJ, 444, 231 (NY95a)

\bibitem[Narayan \& Yi(1995b)]{NY95b}
Narayan, R., \& Yi, I. 1995b, ApJ, 452, 710 (NY95b)

\bibitem[Narayan et al.(1995)]{nar95}
Narayan, R., Yi, I., \& Mahadevan, R. 1995, Nature, 374, 623

\bibitem[Nemmen et al.(2006)]{nem06}
Nemmen, R. S., et al. 2006, ApJ, 643, 652

\bibitem[Noble et al.(2007)]{non07}
Noble, S. C., Leung, P. K., Gammie, C. F., \& Book, L. G.
2007, Class. Quant. Grav., 24, S259

\bibitem[Novikov \& Thorne(1973)]{nt73}
Novikov, I. D., \& Thorne, K. S. 1973, in Blackholes, ed. C. DeWitt,
\& B. DeWitt (Gordon \& Breach) p343

\bibitem[Orosz et al.(1997)]{oro97}
Orosz, J. A., Remillard, R. A., Bailyn, C. D., \& McClitnock, J. E.
1997, ApJ, 478, L83

\bibitem[Orosz et al.(2007)]{oro07}
Orosz, J. A., et al. 2007, Nature, 449, 872

\bibitem[Paczy\'nski(1998)]{pac98}
Paczy\'nski, B. 1998, Acta Astronomica, 48, 667

\bibitem[Pen et al.(2003)]{pen03}
Pen, U. L., Matzner, C. D., \& Wong, S. 2003, ApJ, 596, L207

\bibitem[Perna et al.(2003)]{per03}
Perna, R., Narayan, R., Rybicki, G., Stella, L., \& Treves, A.
2003, ApJ, 594, 936

\bibitem[Piran(1978)]{pir78}
Piran, T. 1978, ApJ, 221, 652

\bibitem[Popham \& Gammie(1998)]{pop98}
Popham, R. G., \& Gammie, C. F. 1998, ApJ, 504, 419

\bibitem[Popham \& Narayan(1991)]{pop91}
Popham, R. G., \& Narayan, R. 1991, ApJ, 370, 604

\bibitem[Quataert(2001)]{qua01}
Quataert, E. 2001, in Probing the Physics of Active Galactic Nuclei by 
Multiwavelength Monitoring, ed. B. M. Peterson, R. S. Polidan \& R. W.
Pogge (Astr. Soc. Pacific) p71

\bibitem[Quataert(1998)]{qua98}
Quataert, E. 1998, ApJ, 500, 978

\bibitem[Quataert et al.(1999)]{q99}
Quataert, E., Di Matteo, T., Narayan, R., \& Ho, L. C. 1999,
ApJ, 525, L89

\bibitem[Quataert \& Gruzinov(1999)]{qg99}
Quataert, E., \& Gruzinov, A. 1999, ApJ, 520, 248

\bibitem[Quataert \& Gruzinov(1999)]{qg00}
Quataert, E., \& Gruzinov, A. 2000, ApJ, 539, 809

\bibitem[Quataert \& Narayan(1999)]{qn99}
Quataert, E., \& Narayan, R. 1999, ApJ, 520, 298

\bibitem[Ramadevi \& Seetha(2007)]{rs07}
Ramadevi, M. C., \& Seetha, S. 2007, MNRAS, 378, 182

\bibitem[Rees et al.(1982)]{ree82}
Rees, M. J., Begelman, M. C., Blandford, R. D., \& Phinney, E. S.
1982, Nature, 295, 17

\bibitem[Remillard et al.(2006)]{rem06}
Remillard, R. A., Lin, D., Cooper, R. L., \& Narayan, R. 2006,
ApJ, 646, 407

\bibitem[Reynolds et al.(1996)]{rey96}
Reynolds, C. S., Di Matteo, T., Fabian, A. C., Hwang, U., \& Canizares,
C. R. 1996, MNRAS, 283, L111

\bibitem[Rozanska \& Czerny(2000)]{rc00}
Rozanska, A., \& Czerny, B. 2000, A\&A, 360, 1170

\bibitem[Rykoff et al.(2007)]{ryk07}
Rykoff, E. S., Miller, J. M., Steeghs, D., \& Torres, M. A. P. 2007,
ApJ, 666, 1129

\bibitem[Sch\"odel et al.(2002)]{sch02}
Sch\"odel, R., et al. 2002, Nature, 419, 694

\bibitem[Serabyn et al.(1997)]{ser97}
Serabyn, E., Carlstrom, J., Lay, O., Lis, D. C., Hunter, T. R., 
\& Lacy, J. H. 1997, ApJ, 490, L77

\bibitem[Shahbaz et al.(2005)]{sha05}
Shahbaz, T., et al. 2005, MNRAS, 362, 975

\bibitem[Shakura \& Sunyaev(1973)]{ss73}
Shakura, N. I., \& Sunyaev, R. A. 1973, A\&A, 24, 337

\bibitem[Shapiro et al.(1976)]{sle76}
Shapiro, S. L., Lightman, A. P., \& Eardley, D. M. 1976, ApJ, 204, 187

\bibitem[Shapiro \& Teukolsky(1983)]{sha83}
Shapiro, S. L., \& Teukolsky, 1983, Black Holes, White Dwarfs, and
Neutron Stars: The Physics of Compact Objects (Wiley-Interscience)

\bibitem[Sharma et al.(2007)]{sha07} 
Sharma, P., Quataert, E., Hammett, G. W., \& Stone, J. M. 2007, ApJ,
667, 714

\bibitem[Sharma et al.(2006)]{sha06}
Sharma, P., Hammett, G. W., Quataert, E., \& Stone, J. M. 2006, ApJ,
637, 952

\bibitem[Shen et al.(2005)]{she05}
Shen, Z. Q., Lo, K. Y., Liang, M. C., Ho, P. T., \& Zhao, J. H.
2005, Nature, 438, 62

\bibitem[Sikora et al.(2007)]{sik07}
Sikora, M., Stawarz, L., \& Lasota, J. P. 2007, ApJ, 658, 815

\bibitem[Smak(1999)]{sma99}
Smak, J. 1999, Acta Astronomica, 49, 391

\bibitem[Spruit \& Deufel(2002)]{sd02}
Spruit, H. C., \& Deufel, B. 2002, A\&A, 387, 918

\bibitem[Spruit et al.(1987)]{spr87}
Spruit, H. C., Matsuda, T., Inoue, M., \& Sawada, K. 1987, 
MNRAS, 229, 517

\bibitem[Stolovy et al.(2003)]{sto03}
Stolovy, S., Melia, F., McCarthy, D., \& Yusef-Zadeh, F. 2003,
Astron. Nachr., 324, 419

\bibitem[Stone \& Pringle(2001)]{sto01}
Stone, J. M., \& Pringle, J. E. 2001, MNRAS, 322, 461

\bibitem[Stone et al.(1999)]{sto99}
Stone, J. M., Pringle, J. E., \& Begelman, M. C. 1999, MNRAS, 310, 1002

\bibitem[Sunyaev \& Titarchuk(1980)]{st80}
Sunyaev, R. A., \& Titarchuk, L. G. 1980, A\&A, 86, 121


\bibitem[Tournear et al.(2003)]{tou03}
Tournear, D., et al. 2003, ApJ, 595, 1058

\bibitem[Ulvestad \& Ho(2001)]{ulv01}
Ulvestad, J. S., \& Ho, L. C. 2001, ApJ, 562, L133

\bibitem[Visser \& Wiltshire(2004)]{vis04}
Visser, M., \& Wiltshire, D. L. 2004, Class. Quantum Grav., 21, 1135

\bibitem[Wu(1997)]{wu97}
Wu, X. 1997, MNRAS, 292, 113

\bibitem[Wu \& Li(1996)]{wu96}
Wu, X., \& Li, Q. 1996, ApJ, 469, 776

\bibitem[Wu et al.(2007)]{wu07}
Wu, Q., Yuan, F., \& Cao, X. 2007, ApJ, 669, 96

\bibitem[Yuan(2001)]{yua01}
Yuan, F. 2001, MNRAS, 324, 119

\bibitem[Yuan(2003)]{yua03}
Yuan, F. 2003, ApJ, 594, L99

\bibitem[Yuan \& Cui(2005)]{yua05}
Yuan, F., \& Cui, W. 2005, ApJ, 629, 408

\bibitem[Yuan et al.(2005)]{ycn05}
Yuan, F., Cui, W., \& Narayan, R. 2005, ApJ, 620, 905

\bibitem[Yuan et al.(2008)]{yua08} 
Yuan, F., Ma, R., \& Narayan, R. 2008, ApJ, in press
(astro-ph/08021679)

\bibitem[Yuan \& Narayan(2004)]{yn04}
Yuan, F., \& Narayan, R. 2004, ApJ, 612, 724

\bibitem[Yuan et al.(2003)]{yqn03}
Yuan, F., Quataert, E., \& Narayan, R. 2003, ApJ, 598, 301

\bibitem[Yuan \& Zdziarski(2004)]{yua04}
Yuan, F., \& Zdziarski, A. 2004, MNRAS, 354, 953

\bibitem[Yuan et al.(2007)]{yua07}
Yuan, F., Zdziarski, A., Xue, Y., \& Wu, X. B. 2007, ApJ, 659, 541

\bibitem[Yungelson et al.(2006)]{yun06}
Yungelson, L. R., et al. 2006, A\&A, 454, 559

\bibitem[Zdziarski \& Gierlinski(2004)]{zg04}
Zdziarski, A. A., \& Gierlinski, M. 2004, Prog. Theor. Phys. Suppl., 
155, 99

\bibitem[Zdziarski et al.(1999)]{zdz99}
Zdziarski, A. A., Lubinski, P., \& Smith, D. A. 1999, MNRAS, 303, L11

\bibitem[Zdziarski et al.(2003)]{zdz03}
Zdziarski, A. A., Lubinski, P., Gilfanov, M., \& Revnivtsev, M. 2003,
MNRAS, 342, 355

\bibitem[Zdziarski et al.(2004)]{zdz04}
Zdziarski, A. A., et al. 2004, MNRAS, 351, 791

\bibitem[Zdziarski et al.(1996)]{zdz96}
Zdziarski, A. A., Johnson, W. N., \& Magdiarz, P. 1996, MNRAS, 283, 193

\bibitem[Zdziarski et al.(1998)]{zdz98}
Zdziarski, A. A., Poutanen, J., Mikolajewska, J., Gierlinski, M., Ebisawa,
K., \& Johnson, W. L. 1998, MNRAS, 301, 435

\bibitem[Zhao et al.(2003)]{zha03}
Zhao, J. H., et al. 2003, ApJ, 586, L29

\bibitem[Zimmerman et al.(2005)]{Zim05}
Zimmerman, E. R., Narayan, R., McClintock, J. E., \& Miller, J. M.
2005, ApJ, 618, 832

\end{thebibliography}
\end{document}